\documentclass[]{interact}

\usepackage{epstopdf}
\usepackage[caption=false]{subfig}

\usepackage[normalem]{ulem}
\usepackage{xcolor}
\usepackage{placeins}
\usepackage{hyperref}

\usepackage[numbers,sort&compress]{natbib}
\bibpunct[, ]{[}{]}{,}{n}{,}{,}
\makeatletter
\def\NAT@def@citea{\def\@citea{\NAT@separator}}
\makeatother

\theoremstyle{plain}

\theoremstyle{definition}

\theoremstyle{remark}

\begin{document}

\articletype{REVIEW ARTICLE}

\title{Quantum Noise Engineering in Advanced LIGO}

\author{
\name{Begüm Kabagöz\thanks{Email: begum@mit.edu}}
\affil{LIGO Laboratory,
Kavli Institute for Astrophysics and Space Research, Massachusetts Institute of Technology, 185 Albany St, Cambridge, MA 02139, USA}
}

\maketitle

\begin{abstract}

\noindent Advanced LIGO is a gravitational-wave detector with extreme sensitivity, operating close to the quantum limits of optical measurement. The quantum noise in this system arises from vacuum fluctuations entering the interferometer readout port: phase-like fluctuations produce measurement imprecision, while amplitude-like fluctuations drive radiation-pressure back-action. In this review, we present an intuitive description in terms of complex fields and quadratures, and follow squeezed vacuum from its generation, through its interaction with the filter cavity and propagation through the interferometer, to the readout. We discuss how frequency-independent and frequency-dependent squeezing engineer quantum noise, and how correlations between optical quadratures map onto correlations between measurement imprecision and back-action. We conclude that LIGO is both a gravitational-wave observatory and a kilometre-scale platform for quantum measurement science.

\end{abstract}


\section{Introduction}
\label{sec:introduction}

Laser Interferometer Gravitational-Wave Observatory (LIGO) interferometers are the most sensitive displacement sensors ever built, sensitive enough to measure the faint spacetime ripples of gravitational waves (GWs) from distant astrophysical collisions~\cite{GW150914}. In Advanced LIGO, two kilometre-scale dual-recycled Fabry--P\'erot Michelson interferometers in Livingston, Louisiana and in Hanford, Washington, measure differential motion of suspended mirror test masses with a precision approaching the quantum limits of optical measurement \cite{AdvancedLIGO2015}. In addition to classical forces that drive test mass motion, we have to consider the light used to measure the mirror displacement as a possible source of disturbance as well. Vacuum fluctuations in the phase quadrature\footnote{Quadratures are the fundamental variables used to describe the quantum state of an electromagnetic field, analogous to position and momentum in quantum mechanics. They represent the in-phase and out-of-phase components of an oscillating field and are particularly useful because quantum measurements and uncertainty relations are expressed naturally in terms of them.} entering the interferometer through the readout port (i.e. the dark port) appear as shot noise in the measurement, while photon-number fluctuations of the amplitude quadrature produce radiation-pressure forces that move the mirrors. These two effects, measurement imprecision and quantum back-action due to the measurement itself, form the central quantum-noise tradeoff of an interferometric position measurement~\cite{Caves1981,BragVor75,Kimble2001Conversion}. The balance between measurement imprecision and its back-action can be expressed quantitatively for any given input laser power as a standard quantum limit~\cite{Caves1980}.

A way to manipulate the effect of this tradeoff on the measurement quadrature we observe is to engineer the quantum state of the light entering the dark port by modifying the incident vacuum state. Injecting squeezed vacuum states into the dark port, first proposed by Caves for interferometric GW detectors~\cite{Caves1981}, has become an operational technology in Advanced LIGO since Observation Run 3 (2019-2020)~\cite{Tse2019QuantumEnhanced, Yu2020QuantumCorrelations}. In the complex-field phase space picture (see e.g. Figs \ref{fig:input_states} and \ref{fig:shearing_nosqz}), ordinary vacuum can be considered as a circular noise cloud centred about the origin (zero mean field) with noise evenly distributed over all optical quadratures. Squeezed vacuum injection replaces this ordinary vacuum state with a state whose fluctuations are redistributed intentionally between quadratures, depicted as an ellipse instead of a circle. If the smaller (lower noise) axis of the ellipse is aligned with the measured signal quadrature, the observed signal-to-noise is improved. 

Notably, the effect of squeezing is not to simply reduce the noise of the injected quantum state in the preferred quadrature. Reducing fluctuations in one quadrature necessarily increases fluctuations in the orthogonal one, as required by the Heisenberg Uncertainty Principle. The noise increase in the orthogonal quadrature matters because an interferometric GW detector is an optomechanical transducer that can couple quadratures: some of the orthogonal quadrature mixes into the observed quadrature. This coupling results in radiation pressure noise. As a result, frequency-independent squeezing reduces high-frequency shot noise while increasing low-frequency radiation-pressure noise. The useful quantity is not only the amount of squeezing generated by the source, but also the orientation of the squeezing ellipse when it reaches the interferometer, and the way the interferometer transforms that ellipse.

Frequency-dependent squeezing addresses this problem by rotating the squeezing ellipse before injection in a frequency-dependent way and leveraging strong correlations that form between the two quadratures of the quantum state. In LIGO, the generated squeezed vacuum is modified using a filter cavity which rotates the squeezing ellipse as a function of measurement frequency so that the quiet quadrature better follows the optomechanical response of the detector. This has enabled broadband quantum-noise reduction in the LIGO detectors, reducing both high-frequency shot noise and the low-frequency penalty associated with radiation pressure simultaneously~\cite{Ganapathy2023Broadband}. 

This paper develops an intuitive field-based description of this feat of quantum engineering. We begin with the complex electric field at the interferometer readout port, and show how the quantum noise couples in to the readout. We discuss the two manifestation of quantum noise: shot noise and radiation pressure noise. We describe frequency-independent squeezing, showing how the squeezed states are generated, how the squeezing ellipse is stabilised, and the effect of frequency-independent squeezing observed in the Advanced LIGO detectors. After this discussion, we present frequency-dependent squeezing results and how the standard quantum limit is surpassed by this technique. 

We then go one step further and follow the transformations and rotations experienced by the injected quantum state throughout the interferometer. The analysis of the quantum noise components using the full multiport interferometer framework reveals the exact mechanism through which radiation pressure back-action is reduced with frequency dependent squeezing. These constitute the toolbox required to truly understand quantum noise engineering in LIGO and how we can do better. 

Finally, we discuss the limitations and future directions of this approach. Optical loss mixes ordinary vacuum back into the squeezed field and rounds the ellipse, while phase noise rotates anti-squeezing into the measured quadrature. Further improvement therefore requires preserving the prepared quantum state through the entire optical system, not only generating more squeezing at the source. We also discuss balanced homodyne detection as the natural next readout-side degree of freedom: a way to freely choose which quadrature of the transformed output field is measured. Together, squeezed-light injection, filter-cavity rotation, and balanced homodyne detection frame LIGO not only as a gravitational-wave observatory, but also as a controllable kilometre-scale quantum measurement system. 

\section{Quantum noise in optical fields}
\label{sec:quantisation}

Before describing the LIGO as a quantum measurement system, we need to establish how quantum fluctuations of an optical field will be expressed throughout this manuscript. An electromagnetic field oscillating at an optical carrier frequency $\omega_0=2\pi\,f_0$ can be described by two components that are $90^\circ$ out of phase with one another. In the quantum description, these components become operators, conventionally called the two optical \textit{quadratures}. We can write the quantised electric field as:
\begin{equation}
    \hat{E}(t) = E_0[\hat{X}_1 \cos(\omega_0t) + \hat{X}_2 \sin(\omega_0t)]
\end{equation}
where $\hat{X}_1$ and $\hat{X}_2$ describe the two orthogonal field quadratures, and $E_0$ is the field normalisation. Geometrically, the two quadratures define perpendicular axes in the complex field plane. A fluctuation of the optical field can therefore be represented by its components along these two directions/quadratures. 

The quadrature description is especially useful because $\hat{X}_1$ and $\hat{X}_2$ are conjugate quantum observables, analogous to position and momentum, rather than two independent classical variables, and as such they obey:

\begin{equation}
    [\hat{X}_1,\hat{X}_2] = \frac{i}{2}, \quad \Delta X_1\Delta X_2 \geq \frac{1}{4}
\label{eq:conjugates}
\end{equation}

\noindent where the deviations $\Delta X_i \equiv \sqrt{\langle(\hat X_i - \langle \hat X_i \rangle)^2 \rangle} $ is the standard deviation, or uncertainty, of the quadrature
\(\hat X_i\) about its mean. The first relation states that the two quadratures do not commute, while the second is the Heisenberg uncertainty relation for the optical field. Their individual fluctuations can be redistributed, but their uncertainties cannot both be made arbitrarily small. This relation is critical for understanding both squeezed light and the quantum measurement tradeoff in an interferometer performing a position measurement. 

\begin{figure}[t!]
\centering
\includegraphics[width=0.8\linewidth]{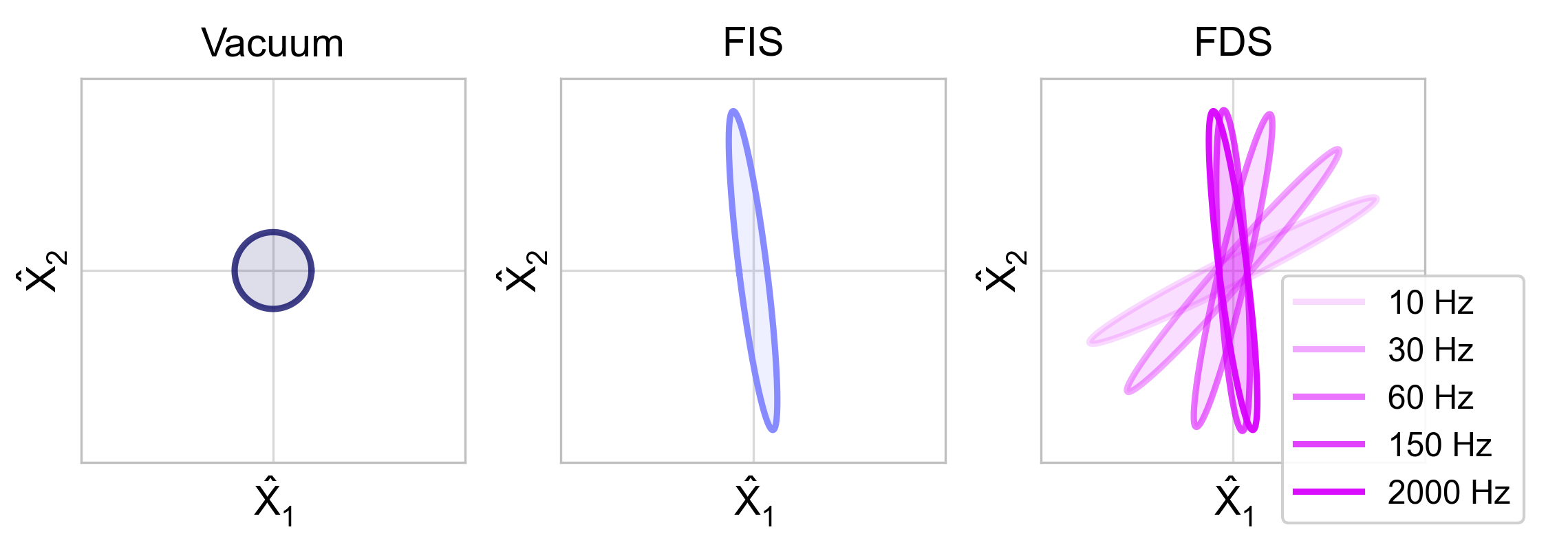}
\caption{
Quadrature representation of the input quantum states considered in this manuscript.
Left: ordinary vacuum, equal fluctuations in all quadrature directions.
Centre: frequency-independent squeezed vacuum (FIS), the uncertainty is reduced along one quadrature and increased along the orthogonal quadrature. The orientation of the ellipse denotes the squeezing angle $\phi$.
Right: frequency-dependent squeezed vacuum (FDS), for which the squeezing angle is a function of frequency, $\phi=\phi(\Omega)$. The ellipse therefore has a different orientation for different frequency components of the quantum field. 
}
\label{fig:input_states}
\end{figure}

The optical vacuum field is an especially important case for us, since it is the origin of quantum noise in the interferometer. While its mean electric field is zero, measurement of either quadrature has a finite statistical spread. Ordinary vacuum has equal uncertainty in the two quadratures and no correlation between them, so its fluctuations are represented by a circular distribution centred at the origin of the quadrature plane, as shown in the left panel of Figure~\ref{fig:input_states}. The circle does not denote a field of fixed amplitude, rather it represents the statistical uncertainty of repeated measurements. In other words, the vacuum state has no preferred readout direction in the quadrature plane, every quadrature has the same variance\footnote{$\operatorname{Var}(X_i)=(\Delta X_i)^2$}.

A \textit{squeezed} state redistributes the uncertainty between the two quadratures. The fluctuations are shrunk below the ordinary vacuum level along one direction and inflated along the orthogonal direction, while remaining bound with Equation~\ref{eq:conjugates}. The circular uncertainty region therefore becomes an ellipse. The minor axis of the ellipse identifies the squeezed quadrature and the major axis the anti-squeezed quadrature. The orientation of the ellipse in the quadrature plane is described by the \textit{squeezing angle}, $\phi$. A representation of the squeezing ellipse is shown in Figure~\ref{fig:input_states} middle panel. Open-source tools for exploring these quadrature representations and simple squeezed-light models are available in the LIGO Quantum Tools repository~\cite{Kabagoz2026LIGOQuantumTools}.

An optical measurement selects a projection of the field in the quadrature space. A measurement performed at quadrature angle $\zeta$ measures

\begin{equation}
    \hat{X}_\zeta = \hat{X}_1\cos\zeta + \hat{X}_2\sin\zeta.
\end{equation}

The angle $\zeta$ is therefore the \textit{measurement} or \textit{readout quadrature} angle. Geometrically, the measurement returns the projection of the field fluctuations onto this direction. The measured quantum noise therefore depends both on the quantum state of the field and on the quadrature selected by the readout. For a squeezed state, rotating the squeezing ellipse changes the fluctuations projected onto a fixed measurement quadrature. Alternatively, changing the readout quadrature changes which part of the same quantum state is observed.

LIGO is, in essence, a continuous optical measurement of the differential position of its test masses. In this measurement, the two optical quadratures are associated with phase-like and amplitude-like fluctuations of the field: uncertainty in the phase-like quadrature limits the precision with which the position can be estimated, while uncertainty in the amplitude-like quadrature couples to the mechanical motion of the masses. The uncertainty relation between the two conjugate quadratures therefore has a direct consequence for the attainable measurement precision. In a conventional position measurement, this relationship gives rise to the standard quantum limit (SQL), which we will discuss in more detail in Section~\ref{sec:shot-rad}.

\section{Laser Interferometer Gravitational Wave Observatory (LIGO)}
\label{sec:background}

Before launching into how the input quantum states interact with the system and generate quantum noise, it is useful to review some key features of the interferometer that will be our quantum measurement platform throughout the paper: the optomechanical system, the meaning of a sensitivity curve, and the sideband description of audio-frequency signals. These concepts provide the bridge between the optical field picture and the measured detector noise.

\begin{figure}[hbt!]
    \centering
    \includegraphics[width=0.65\linewidth]{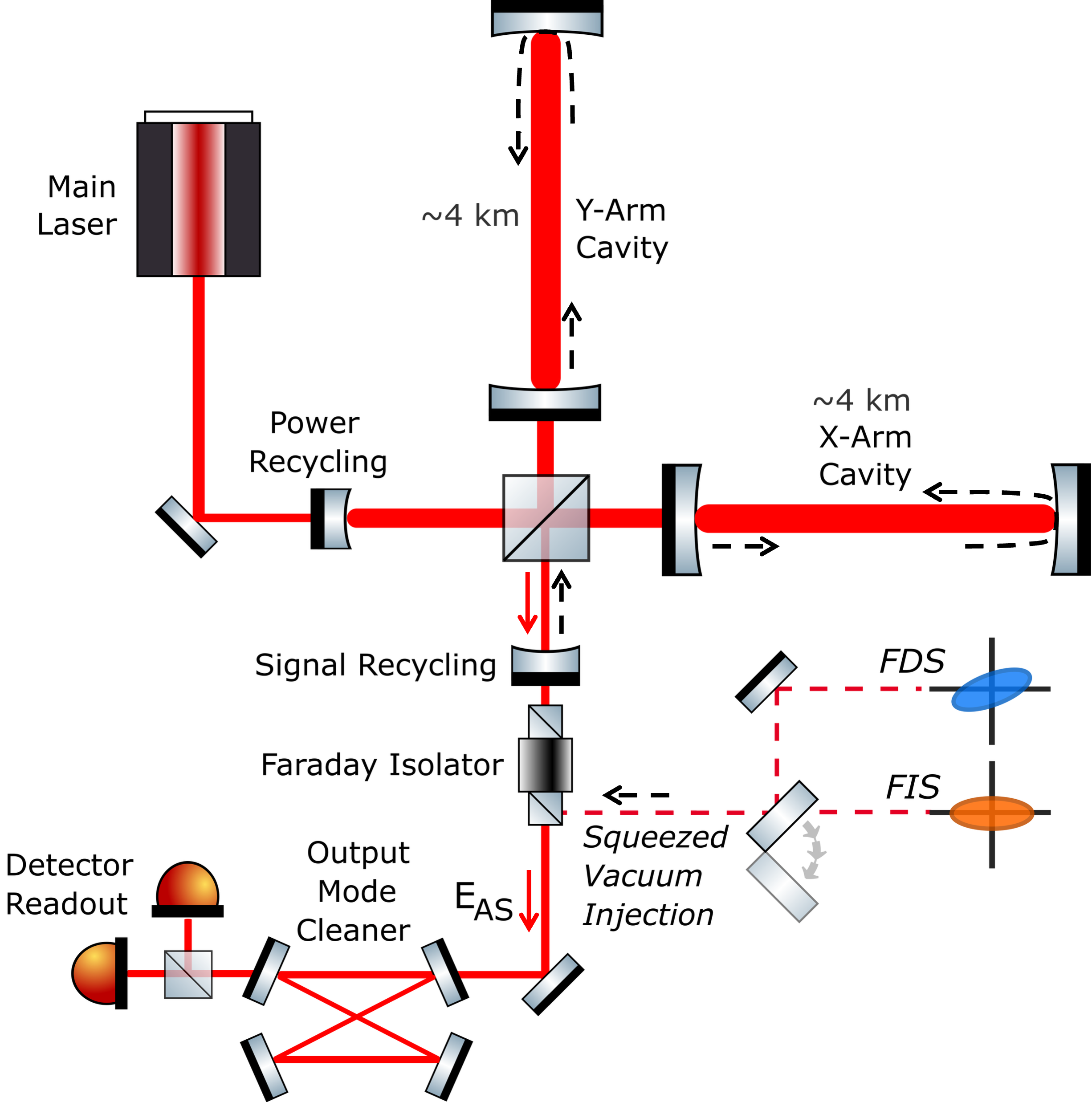}
    \caption{
   Interferometer optical diagram and quantum-field propagation. The red solid path shows the coherent beam propagating within the interferometer. The dashed red path shows squeezed-vacuum injection, the black dashed arrows trace its
   propagation through the interferometer, and the solid red arrow denotes the outgoing field \(E_{\rm AS}\). With no squeezed-vacuum injection, the corresponding incident mode is in the ordinary vacuum state. FIS and FDS denote frequency-independent (see Section~\ref{sec:fis}) and frequency-dependent squeezing (see Section~\ref{sec:fds}), respectively.
}
    \label{fig:ifo}
\end{figure}

Advanced LIGO is a dual-recycled Michelson interferometer, with two 4-km long arms, as shown schematically in Fig. \ref{fig:ifo}. Each arm contains a Fabry-P\'erot cavity comprised of 40-kg mirror test masses suspended as the final stage of a quadruple pendulum~\cite{Cumming2012AdvancedLIGOSuspension}, which gives rise to a mechanical response. The mechanical susceptibility of the test masses $\chi_m(\Omega)$ describes their displacement response to an applied force, $x(\Omega)=\chi_m(\Omega)F(\Omega)$. This frequency-dependent response plays an important role in coupling radiation-pressure force to mechanical motion.

The main laser enters through the "bright port", is split into two beams with the beam splitter and sent to the X and Y arms. An incoming gravitational wave changes the differential length of the two arms as experienced by the photons, producing an optical phase difference between the beams returning from the arms. Each arm contains a resonant cavity that circulates the main beam many times between two 40-kg mirrors, building up the GW imprint as well as the total optical power in the arms during the storage time. The relative phase difference of the beams returning from the arms determines if they constructively or destructively interfere at the ports of the beam splitter. While the port with the main laser incident, dubbed the "bright" port, receives the majority of the returned photons, the output or "dark" port sees nearly complete destructive interference. A power recycling mirror at the input/bright port forms the power recycling cavity, which builds up power circulating in the interferometer. The dark port, where the GW signal is read out, is necessarily more complicated.

The GW signal, imprinted as audio-frequency sidebands on a carrier field, exits primarily through the dark (or antisymmetric) port. There it encounters the signal recycling/extraction cavity, which, together with the arm
cavities, shapes the frequency response of the interferometer. For the
configuration considered here, the coupled-cavity pole is approximately
\(\sim 440\,\mathrm{Hz}\), providing broadband sensitivity over frequencies
important for binary-neutron-star inspirals~\cite{AdvancedLIGO2015}. The outgoing field then passes through the output mode cleaner, which rejects higher-order spatial modes and radio-frequency control sidebands, before reaching the photodetectors.

The same dark port is also the port through which vacuum fluctuations enter the measurement. In the absence of squeezed-light injection, ordinary vacuum enters through the dark port, couples to the interferometer, and is finally detected alongside the signal field \cite{Caves1981}. In the squeezed-light configuration, this ordinary vacuum is replaced by squeezed vacuum. The squeezed vacuum is injected through the reject port of the Output Faraday Isolator\footnote{which is a passive device that operates like the optical equivalent of a diode, stopping back-reflections from coupling back to the interferometer}~\cite{Tse2019QuantumEnhanced,Ganapathy2023Broadband}. The black dashed arrows in Figure~\ref{fig:ifo} indicate the injection path of squeezed-light, and the subsequent propagation through the interferometer. The solid red arrow indicates the total output field at the dark port, including both the coherent beam from the interferometer and the (ordinary or squeezed) vacuum that travels alongside it. Thus, the dark port plays two roles at once. It is the output port for the gravitational-wave signal, and it is the input port for the quantum noise that limits the readout.

\FloatBarrier

The performance of a GW detector is usually characterised by a sensitivity curve, which shows the smallest GW strain that could be distinguished from detector noise at each frequency. The sensitivity is plotted as an amplitude spectral density, usually in units of \(1/\sqrt{\rm Hz}\) for strain\footnote{The gravitational-wave strain \(h\) is the fractional differential change $\Delta L$ in arm length $L$, $ h = \Delta L/L$.} or \({\rm m}/\sqrt{\rm Hz}\) for equivalent displacement. Lower noise curves mean better sensitivity: a smaller strain or displacement would be visible above the noise.

A measured sensitivity curve is the total result of many noise sources. At low frequencies, seismic motion, suspension thermal noise, angular control noise, and other technical couplings can be important. In the band most sensitive to desired astrophysical signals of interest\footnote{such as Binary Neutron Stars, whose sensitivity band peaks around 50 Hz}, coating thermal noise and other classical displacement noises contribute strongly. At higher frequencies, photon shot noise becomes increasingly important. When the power circulating in the interferometer is high, radiation-pressure noise dominates low frequencies (due to the mechanical susceptibility).

\begin{figure}[t!]
    \centering
    \includegraphics[width=0.8\linewidth]{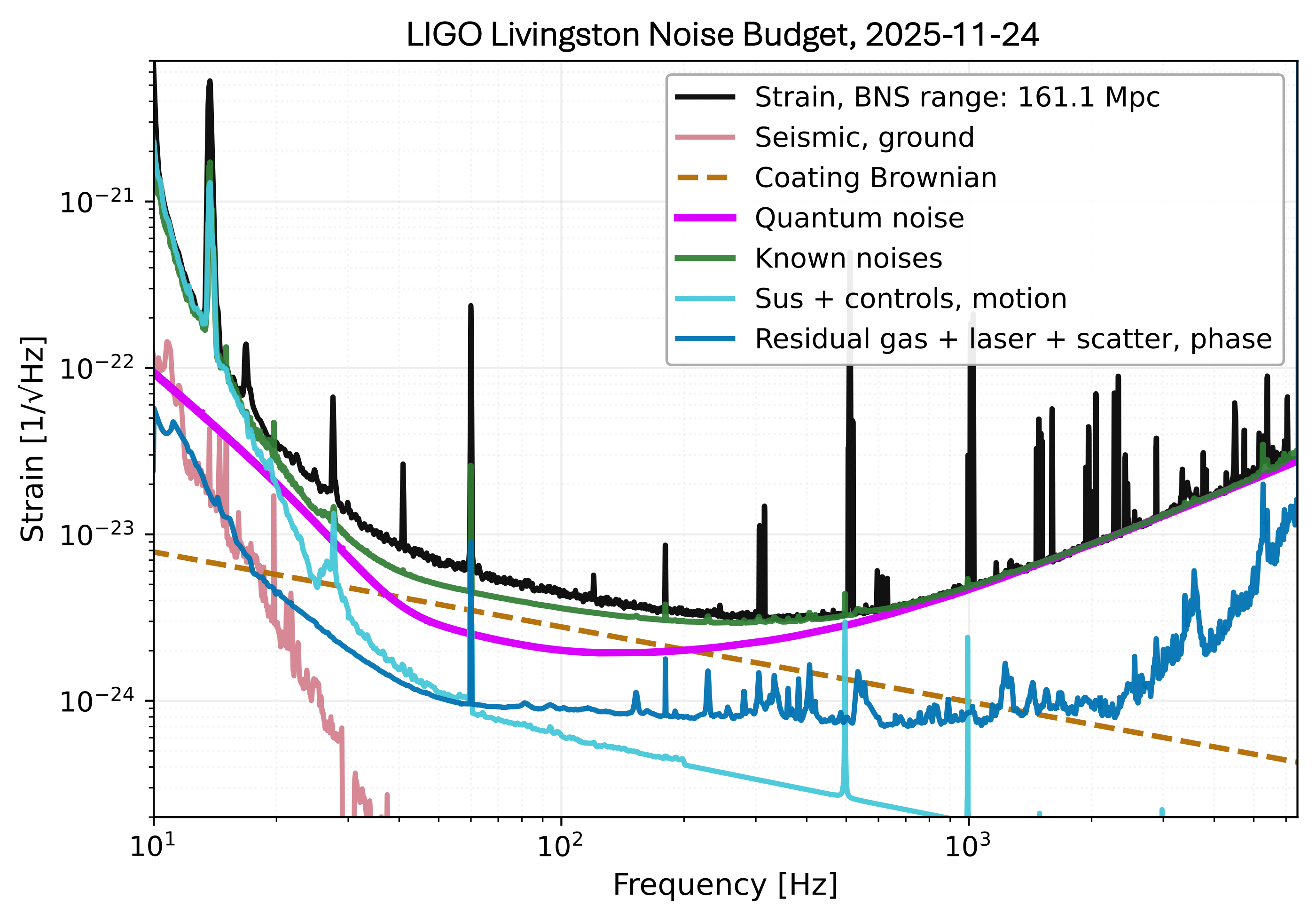}
    \caption{
    Example LIGO Livingston sensitivity curve and noise budget.
    The measured strain sensitivity is the total noise of the detector, not only quantum noise.
    Quantum-noise reduction is visible only in frequency bands where other classical noise sources have been sufficiently suppressed.
    This motivates the system-level view used in this paper: squeezed light improves the detector only when it is delivered through an interferometer quiet enough to reveal the quantum-noise contribution. 
    }
    \label{fig:llo_noise_budget}
\end{figure}

Figure~\ref{fig:llo_noise_budget} shows the noise budget and sensitivity curve for LIGO Livingston Observatory on a representative time and date for Observing Run 4 (2023-2025). Notice how quantum noise (bold solid magenta line) is a prominent source of noise, with many classical noise sources suppressed below it. Squeezing can improve the detector only in frequency bands where quantum noise is a significant part of the total noise. The squeezed state must be generated and injected, but the rest of the interferometer must also be quiet enough for the quantum-noise reduction to make a difference. The broader detector noise context is reviewed in Ref.~\cite{McIver2020Discovering}, discussed in detail for Observing Run 4 in Ref.~\cite{advanced_capote_2025}; here we focus on the quantum part of the noise budget and how it can be engineered.
\FloatBarrier

It is useful here to briefly introduce the sideband picture used throughout the rest of the manuscript. A modulation of the optical carrier ($\omega_0$) at an audio frequency $\Omega$, produced by a gravitational wave for instance, appears as upper and lower optical sidebands at $\omega_0+\Omega$ and $\omega_0-\Omega$. Quantum fluctuations at the same audio frequency are likewise represented by this pair of sidebands. The amplitude and phase quadratures introduced in Section~\ref{sec:quantisation} are constructed from combinations of these upper and lower sidebands\cite{CavesSchumaker1985}. When we refer to the quantum state or squeezing angle at a specific measurement frequency, we are describing the fluctuations of the sideband pair at $\omega_0\pm\Omega$.

Often, the quantum noise reduction compared to the unmodified vacuum case is quantified with a "dB squeezing level". If \(V\) is a quadrature variance normalised such that ordinary vacuum has \(V=1\), then
    ${\rm sqz~level~(dB)}
    =
    10\log_{10} V$.
A noise level of \(-3\,{\rm dB}\) therefore means that the variance, or power spectral density, has been reduced by a factor of two. On an amplitude spectral density plot, this corresponds to a reduction by a factor of \(\sqrt{2}\).

\section{Quantum noise and squeezed-light engineering}
\label{sec:qnoise_intuition}

\subsection{How does the quantum noise couple into the interferometer?}

The complex optical field at the antisymmetric port can be written as
$\mathcal{E}_{\rm AS}(t)
    =
    E_{\rm AS}(t)e^{-i\omega_0 t}
    +
    {\rm c.c.}$\footnote{complex conjugate},
where \(E_{\rm AS}(t)\)\footnote{see Figure~\ref{fig:ifo}, the field entering the readout chain, which is comprised of the output mode cleaner and the photodetectors} is the slowly varying complex amplitude and \(\omega_0\) is the carrier frequency. 
The beam splitter combines the fields coming from the X and Y arms. No coherent carrier light is intentionally injected from the antisymmetric port. In the quantum description, however, an open optical port is not simply empty: it sources the vacuum state, whose fluctuations can couple to the interferometer~\cite{Caves1981}. Therefore, the field at the readout may be written as

\begin{equation}
E_{\rm AS}(t)
=
{E_x(t)e^{i\theta_x(t)}-E_y(t)e^{i\theta_y(t)}+{E}_{\rm q}(t).}
\label{eq:eas}
\end{equation}
Here \(E_x(t)\) and \(E_y(t)\) are the real, slowly varying amplitudes
of the fields returning from the X and Y arms to recombine at the beam splitter, while \(\theta_x(t)\) and \(\theta_y(t)\) are the phases acquired by these fields. The minus sign follows from the antisymmetric-port convention. These fields include the coherent carrier light, the signal sidebands produced by differential arm motion, and classical fluctuations that enter through the interferometer. 

Here the $E_{\rm q}$ term denotes the quantum contribution. Although
the underlying quantum field is described formally by an operator, for
the geometric discussion here we represent its fluctuations using
ordinary complex phasors. Accordingly, \(E_{\rm q}(t)\) is written
without a hat and represents a possible instantaneous realisation of
the quantum-field fluctuations. In the absence of squeezed-light injection this contribution corresponds to ordinary vacuum fluctuations; with squeezing, the statistics of these fluctuations are modified, represented with ellipses in Figure \ref{fig:input_states}. 

This notation is useful because all coherent carrier light at the readout port comes from the combination of $E_x$ and $E_y$. In an ideal Michelson interferometer held exactly on the dark fringe, these two fields would destructively interfere perfectly. In the real detector, a small residual carrier remains because of the intentional offset used for DC readout~\cite{Fricke2012DCReadout} and because of imperfect spatial overlap of the two recombining fields.

In DC readout, the interferometer is intentionally held slightly away from the minimum of the dark fringe, by actuating on the test masses to have a constant differential displacement between the arms~\cite{Fricke2012DCReadout}. Let's call the resultant controlled interferometer field \(E_{\rm IFO}\), as in Figure \ref{fig:ifo}. The interferometer control system keeps the interferometer close to this operating point, where small differential arm displacements produce approximately linear changes in the detected optical power. The feedback system acts on the observed motion, made up of gravitational-wave signals, residual classical motion, and radiation-pressure-driven motion, which are ultimately represented in the measured sensitivity curve.

For the purposes of understanding the quantum sensing problem here, let us pose the readout in terms of the quantum measurement. We can express the field that reaches the readout as $E_{\rm AS}(t)=E_{\rm IFO}(t)+E_{\rm q}(t)$, ignoring the classical fluctuations that would normally be contained in the dark port coherent field. The role of \(E_{\rm IFO}\) is to act as the local oscillator: the phase reference against which the quantum field beats at the photodetector. Here $E_{\rm q}(t)$ denotes an additive quantum-field contribution, rather than a fluctuation of the coherent interferometer field $E_{\rm IFO}$. The photodetector measures optical power which is proportional to the intensity of the field received,
\begin{align}
P_{\rm{det}}(t)
&\propto
\left|
E_{\rm IFO}
+
E_{\rm q}(t)
\right|^2
\nonumber \\
&=
|E_{\rm IFO}|^2
+
2{\rm Re}
\left[
E_{\rm IFO}^*E_{\rm q}(t)
\right]
+
|E_{\rm q}(t)|^2 .
\label{eq:readoutP}
\end{align}
The first term is the optical power associated with the coherent interferometer field. The second term is the linear beating/interference of the quantum field against this coherent field. Because \(|E_{\rm IFO}|\gg|E_{\rm q}|\), the final quadratic term is negligible. Strictly speaking, assigning a classical phase to vacuum is not appropriate. For building geometric intuition however, we can imagine an instantaneous fluctuation as a small phasor $E_q$. The quantum contribution to the detected photodetector signal therefore becomes:
\begin{equation}
\delta P_{\rm q}(t)
\propto
2{\rm Re}
\left[
E_{\rm IFO}^* E_{\rm q}(t)
\right] = 2|E_{\rm{IFO}}|\,|E_{\rm{q}}|\,\cos(\zeta-\theta_q).
\label{eq:quantum_beating}
\end{equation}

This expression is the key: \(E_{\rm IFO}\) provides the coherent reference field against which small differential signals are read out, while the $\delta P_{\rm q}$ term represents the quantum fluctuation that limits the precision of this measurement. At the photodetector this noise contribution appears as imprecision noise, or shot noise. The photodetector does not measure the full complex quantum field; rather, it measures the projection of that field along the phasor of \(E_{\rm IFO}\), which is represented by the factor $\cos(\zeta-\theta_q)$, where $\zeta$ and $\theta_q$ are phases associated with \(E_{\rm IFO}\) and \(E_{\rm q}\), respectively.

The phase of \(E_{\rm IFO}\) selects the readout quadrature of the interferometer and, therefore, which quadrature of quantum field is detected,  that is, $\zeta$ is the readout angle. 
Geometrically, vacuum fluctuations are represented by a circular distribution in the complex field plane, as seen in Figure~\ref{fig:input_states}.
Because the distribution is rotationally symmetric, there is no preferred phase or quadrature direction: fluctuations are equally likely to project along any direction in the complex plane, and every quadrature has the same statistical spread. Consequently, no choice of the interferometer phase readout can select a quieter quadrature.

Equation~\ref{eq:quantum_beating} describes the quantum fluctuation in detected-power units, but we need to refer this fluctuation to an equivalent differential arm displacement to make sense of the quantum noise in strain sensitivity units. A small differential displacement \(x(\Omega)\) changes the relative phase of the arm fields and produces a signal sideband whose amplitude scales as $E_{\rm sig}(\Omega)\propto E_{\rm arm}\,x(\Omega),$ where \(E_{\rm arm}\) represents the carrier field circulating in the arms and the frequency-dependent optical response has been ignored here for clarity. The beating of this signal field with \(E_{\rm IFO}\) gives an optical response satisfying $
\left|
\frac{\partial P_{\rm det}}{\partial x}
\right|
\propto
|E_{\rm IFO}|\,|E_{\rm arm}|.
$

The equivalent displacement imprecision therefore scales as

\begin{equation}
\delta x_{\rm imp}
\sim
\frac{\delta P_{\rm q}}
{\left|\partial P_{\rm det}/\partial x\right|}
\propto
\frac{|E_{\rm IFO}|\,|E_{\rm q}|}
{|E_{\rm IFO}|\,|E_{\rm arm}|}
=
\frac{|E_{\rm q}|}{|E_{\rm arm}|}.
\label{eq:shot noise dep on parm}
\end{equation}

Because \(P_{\rm arm}\propto |E_{\rm arm}|^2\), the corresponding amplitude and power spectral densities scale as $\sqrt{S_{xx}^{\rm imp}}\propto1/{\sqrt{P_{\rm arm}}}, \,\,
S_{xx}^{\rm imp}\propto1/{P_{\rm arm}}
$, yielding the familiar inverse dependence of quantum shot noise with arm power.

\subsection{Shot noise and Radiation-pressure noise}
\label{sec:shot-rad}

The field-level discussion above explains how vacuum fluctuations appear as quantum shot noise at the readout. This noise limits how well we can measure the interferometer field, and therefore how well we can infer the differential motion of the mirrors. 

\begin{figure}[hbt!]
    \centering
    \includegraphics[width=0.45\linewidth]{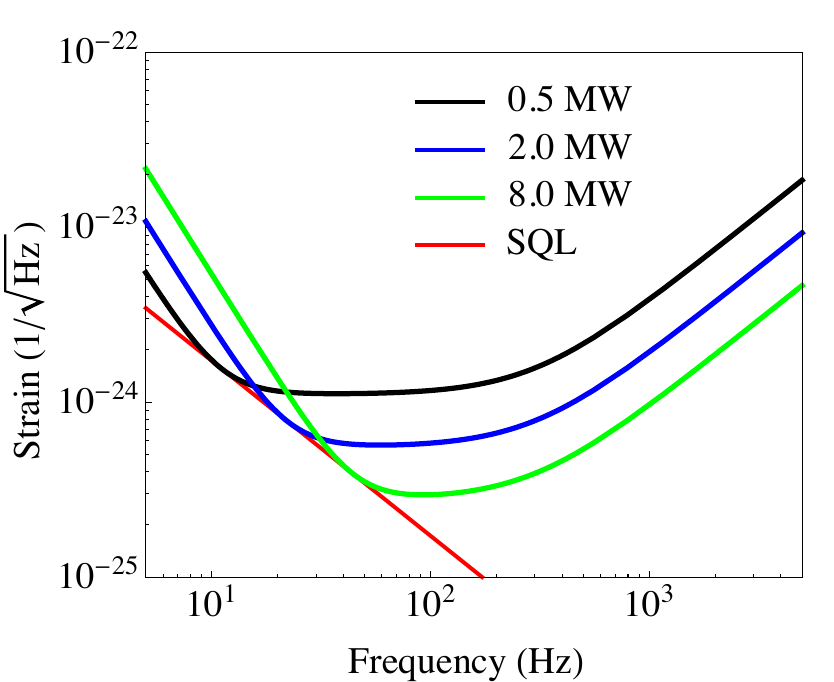}
    \caption{
Quantum noise due to unmodified vacuum with increasing arm power.
Reproduced from Fig.~7 of~\cite{DanilishinKhaliliMiao2019}
© The Author(s) 2019, licensed under CC BY 4.0.}
    \label{fig:shot_rad_power}
\end{figure}

If shot noise, or \textbf{imprecision}, were the only quantum effect, the way to reduce quantum noise experienced by the system would be simple: increasing the signal level by increasing laser power. A larger circulating arm field produces a larger beat signal for the same mirror displacement, so the vacuum fluctuations \(E_{\rm q}\) corresponds to a smaller equivalent displacement noise, see Equation~\ref{eq:shot noise dep on parm}. This is why the displacement-referred  shot noise decreases with optical power. Figure \ref{fig:shot_rad_power} shows simulated quantum noise curves for another gravitational wave interferometer design rather than Advanced LIGO at three circulating arm powers, showing analogous behaviour~\cite{DanilishinKhaliliMiao2019}. Above 50 Hz, the quantum noise is predominantly shot noise, and the power dependence is clearly seen. At low frequencies, however, the power dependent behaviour is suddenly reversed. This is because shot noise is not the only manifestation of quantum noise in this system. 

The other quantum noise component is called radiation pressure noise. While shot noise is mostly associated with uncertainty in the phase quadrature of the quantum field, radiation pressure noise is associated with uncertainty in the amplitude quadrature. 
The photons used to measure the mirror displacement also appreciably push on the mirrors. A photon carries momentum of \(\hbar k\) and upon reflection from a mirror, it transfers approximately $\Delta p \simeq 2\hbar k$
of momentum to the mirror. This force exerted by light is called radiation-pressure\footnote{radiation pressure is the basis for concepts such as solar-sails}. In LIGO, it acts on the suspended test masses of the arm cavities. The amplitude fluctuations of the quantum field cause fluctuations in the photon flux circulating in the arms, and therefore in the radiation-pressure force applied to the mirrors. If the optical power were perfectly steady, the radiation-pressure force would also be steady and could be treated as part of the operating point. Instead, quantum fluctuations produce a fluctuating force on the test masses.

\FloatBarrier

Radiation-pressure noise is the \textbf{back-action} of the quantum measurement. The same light used to determine the mirror position also exerts a fluctuating force on the mirrors, changing the quantity being measured. Increasing the optical power therefore has opposite consequences for the two quantum-noise manifestations. It improves the measurement precision, reducing shot-noise imprecision, but at the same time increases the photon flux and hence the fluctuating radiation-pressure force acting on the test masses. Radiation-pressure noise therefore increases with optical power, as seen at low frequencies in Figure~\ref{fig:shot_rad_power}.

This gives rise to the central quantum measurement tradeoff in an interferometer: one cannot simply increase the optical power and improve the sensitivity across the entire frequency range. The most general form of total quantum noise comprises of three elements: shot noise, radiation pressure noise and their cross-correlation, respectively:
\begin{equation}
    S_{xx}^{\rm QN} = S_{xx}^{imp} + |\chi_m|^2S_{FF}^{ba} + 2Re[\chi_m^*S_{xF}].
\label{eq:full_qnoise}
\end{equation}

with $\chi_m$ denoting the mechanical susceptibility of the system. 
At a given frequency, increasing power P reduces the imprecision noise but increases the measurement back-action. For a conventional position measurement in which these two noise contributions are uncorrelated,
\begin{equation}
S_{xx}^{\rm imp} \propto \frac{1}{P_{arm}},
\qquad
S_{FF}^{\rm ba} \propto P_{arm},
\label{eq:power relation}
\end{equation}

\noindent where $S_{xx}^{\rm imp}$ describes the displacement imprecision and $S_{FF}^{\rm ba}$ the fluctuating back-action force, with $P_{arm}$ as the arm power. The latter produces displacement noise through the mechanical response of the mirrors. Because detector sensitivity is expressed as equivalent displacement noise, it is useful to refer the latter two contributions to displacement:
\begin{equation}
S_{xx}^{\rm ba}
\equiv
|\chi_m|^2S_{FF}^{\rm ba},
\qquad
S_{xx}^{\rm imp,ba}
\equiv
2\,\mathrm{Re}\!\left[\chi_m^*S_{xF}\right].
\label{eq:displacement_referred_terms}
\end{equation}
The total quantum noise can therefore be written compactly as
\begin{equation}
S_{xx}^{\rm QN}
=
S_{xx}^{\rm imp}
+
S_{xx}^{\rm ba}
+
S_{xx}^{\rm imp,ba}.
\label{eq:qnoise_decomposition}
\end{equation} 

Going back to Equation~\ref{eq:power relation}, there is, therefore, an optical power at which the total quantum noise at a given frequency is minimised: below this power the measurement is limited primarily by imprecision, while above it the increasing back-action becomes dominant. For an ideal continuous position measurement with uncorrelated imprecision and back-action, this minimum defines the \textbf{standard quantum limit (SQL)}. The red trace in Figure~\ref{fig:shot_rad_power} shows this limit. This tradeoff becomes relevant to the measured detector sensitivity only in frequency regions where quantum noise is comparable to or larger than the classical noise sources. Note that the SQL is frequency dependent, and that the power minimising the total quantum noise differs at every frequency. No single interferometer configuration therefore sits at the standard quantum limit across the band; the red trace is the envelope of these per-frequency optima rather than a curve that any one detector follows.

\subsection{Frequency-Independent Squeezing (FIS)}

Squeezing is a technique for shaping the injected quantum state, and hence the quantum noise experienced by the interferometer. The squeezer prepares the vacuum field entering the dark port so that its fluctuations are redistributed between two orthogonal quadratures in a favourable way. In the phasor, or quadrature, picture introduced in Section~\ref{sec:quantisation},
ordinary vacuum is a circular noise cloud. Squeezed vacuum is instead represented by an ellipse: the fluctuations are reduced along one quadrature and increased along the orthogonal quadrature, as shown in the middle panel of Figure~\ref{fig:input_states}. The quantum fluctuations are therefore said to be ``squeezed.''

In the early years of Advanced LIGO, the optical power was not yet large enough for radiation-pressure noise to become a major noise source relative to classical noise at low frequencies. The first applications of squeezing therefore focused on reducing shot-noise imprecision. This approach is known as \textit{frequency-independent squeezing} (FIS), because the same squeezed quadrature is injected across the entire detection band. 

Audio-band squeezing was a technological challenge that needed to be resolved for its use-case in GW detectors, and the first table-top demonstration was in 2004~\cite{McKenzie2004}. Phase stabilization using a co-propagating coherent field was developed in 2006~\cite{CLF}. These developments and their relevance to GW detection are reviewed
in~\cite{Schnabel2010}. The first application of squeezed light in a gravitational-wave observatory was demonstrated at GEO\,600~\cite{Abadie2011GEO600Squeezing}. A proof of principle demonstration was done in LIGO in 2013~\cite{Aasi2013Squeezing}. In 2019, during Observing Run 3, for the first time ever, squeezed quantum states were used in direct measurement of gravitational waves, improving the sensitivity of LIGO above 50 Hz by approximately 3 dB~\cite{Tse2019QuantumEnhanced}.

\subsubsection{Generation of Squeezed States}
\label{sec:squeezing_generation}

\begin{figure}[hbt]
    \centering
    \includegraphics[width=0.7\linewidth]{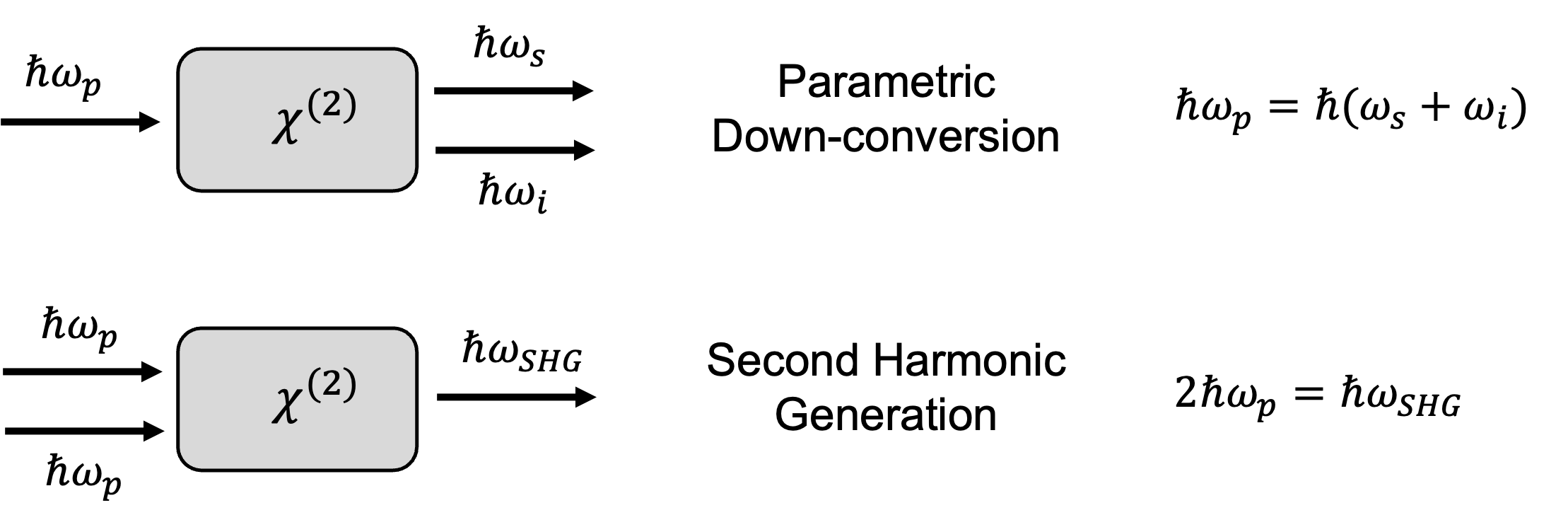}
    \caption{
Diagrams for two $\chi^{(2)}$ (second order nonlinear) processes, parametric down-conversion and second harmonic generation used in LIGO squeezing system. Second harmonic generation is the reverse process to wavelength-degenerate parametric down-conversion. Notice pump frequency $\omega_p$ correspond to different frequencies for the two processes. The term pump denotes the field driving each nonlinear process and therefore corresponds to different frequencies in the two diagrams.
    }
    \label{fig:chi_processes}
\end{figure}

The device that prepares the engineered quantum state is an optical parametric oscillator (OPO), operated below its threshold. An OPO is an optical cavity containing a nonlinear $\chi^{(2)}$ crystal, which receives a "pump" photon with frequency $\omega_p$ and generates two photons with frequencies $\omega_s$ (signal) and $\omega_i$ (idler) from it, see Figure~\ref{fig:chi_processes}.
While the purpose of the cavity is to recirculate the optical fields to improve the efficiency of the conversion process of the desired mode\footnote{longitudinal and spatial mode}, the nonlinear crystal provides the medium in which the nonlinear interaction between the pump, signal and idler photons take place.

Of course, conservation of energy applies: $\hbar \omega_p = \hbar \omega_s+\hbar \omega_i$, and since the resultant photons each have lower energy than the pump photon, this process is also called parametric down-conversion. For efficient conversion, momentum conservation is also needed\footnote{Nonlinear conversion can happen with imperfect phase-matching, but diminished compared to the phase-matched case.}: $\hbar k_p - \hbar k_s - \hbar k_i + \hbar \Delta k \simeq 0$ . This is called the phase matching condition, and $\Delta k$ is a term that can be tuned with periodic poling\footnote{Periodic poling is a technique where the ferro-electric domains of crystals are flipped periodically, essentially creating a phase matching grating. The quasi-phase matching term is related to the poling period $\Lambda$ with $\Delta k(\rm T) = 2\pi/\Lambda(\rm T) $, dependent on temperature $\rm T$.} and temperature control of the crystal. In LIGO, a periodically poled potassium titanyl phosphate (PPKTP) crystal is mounted inside a bowtie cavity, with an oven for temperature control, which is a fairly standard configuration for OPOs. In the degenerate configuration used for LIGO squeezing, the down-converted field is centred at 1064nm, half the pump frequency and the same wavelength as the interferometer carrier.

Now that we established how the photon pairs are generated, let us consider how this nonlinear interaction produces the squeezed-vacuum ellipse.
By virtue of the nonlinear process, the phases of the signal and idler fields are related to the pump phase: $\phi_s+\phi_i = \phi_p+\phi_{NL}$, where $\phi_{NL}$ is the phase associated with the nonlinear interaction. In squeezed vacuum generation, vacuum fluctuations seed the process. The individual phases of the signal and idler fields are therefore not fixed independently; rather, their phase sum is constrained relative to the pump phase\footnote{The individual phases being undetermined while the phase of the entangled photon pairs being determined provides a resource for quantum cryptography~\cite{qcrypto}.}. It follows that, within each generated pair, the phases associated with the signal and idler photons would be correlated with each other. In the degenerate case, the signal and idler photons occupy upper and lower sidebands $\omega_0\pm\Omega$, and together they form the squeezing ellipse at measurement frequency $\Omega$. Their pairwise phase correlations appear as phase-sensitive correlations in the fluctuations at measurement frequency $\Omega$, with the pump phase setting the orientation of the squeezing ellipse quadratures. 
The OPO is therefore a phase-sensitive amplifier: fluctuations along one quadrature are amplified, while fluctuations along the perpendicular quadrature are deamplified. The uncertainty is not reduced unconditionally; it is redistributed between quadratures. The quadrature whose uncertainty is reduced is said to be squeezed, while the quadrature whose uncertainty is increased is anti-squeezed. Access to squeezed vacuum then becomes a matter of aligning the squeezed quadrature with the desired readout quadrature.

\subsubsection{Phase Control of Squeezed States}

The interferometer only benefits from squeezing if the quiet quadrature is aligned with the quadrature being read-out. If the relative phase drifts/fluctuates, the squeezing ellipse rotates/jitters with respect to the interferometer field. The detector then sees a mixture of the squeezed and anti-squeezed quadratures, reducing the observed squeezing and even injecting mostly anti-squeezing in some cases. This is why, for effective use of squeezing, one has to have a method to control the phase of the squeezing ellipse. Squeezed vacuum has no coherent carrier from which its phase can be measured directly. An ancillary field that co-propagates in the OPO, called the coherent locking field (CLF), provides a coherent proxy whose phase is tied to the same nonlinear interaction that determines the orientation of the squeezing ellipse~\cite{CLF}.

\begin{figure}[hbt]
    \centering
    \includegraphics[width=0.9\linewidth]{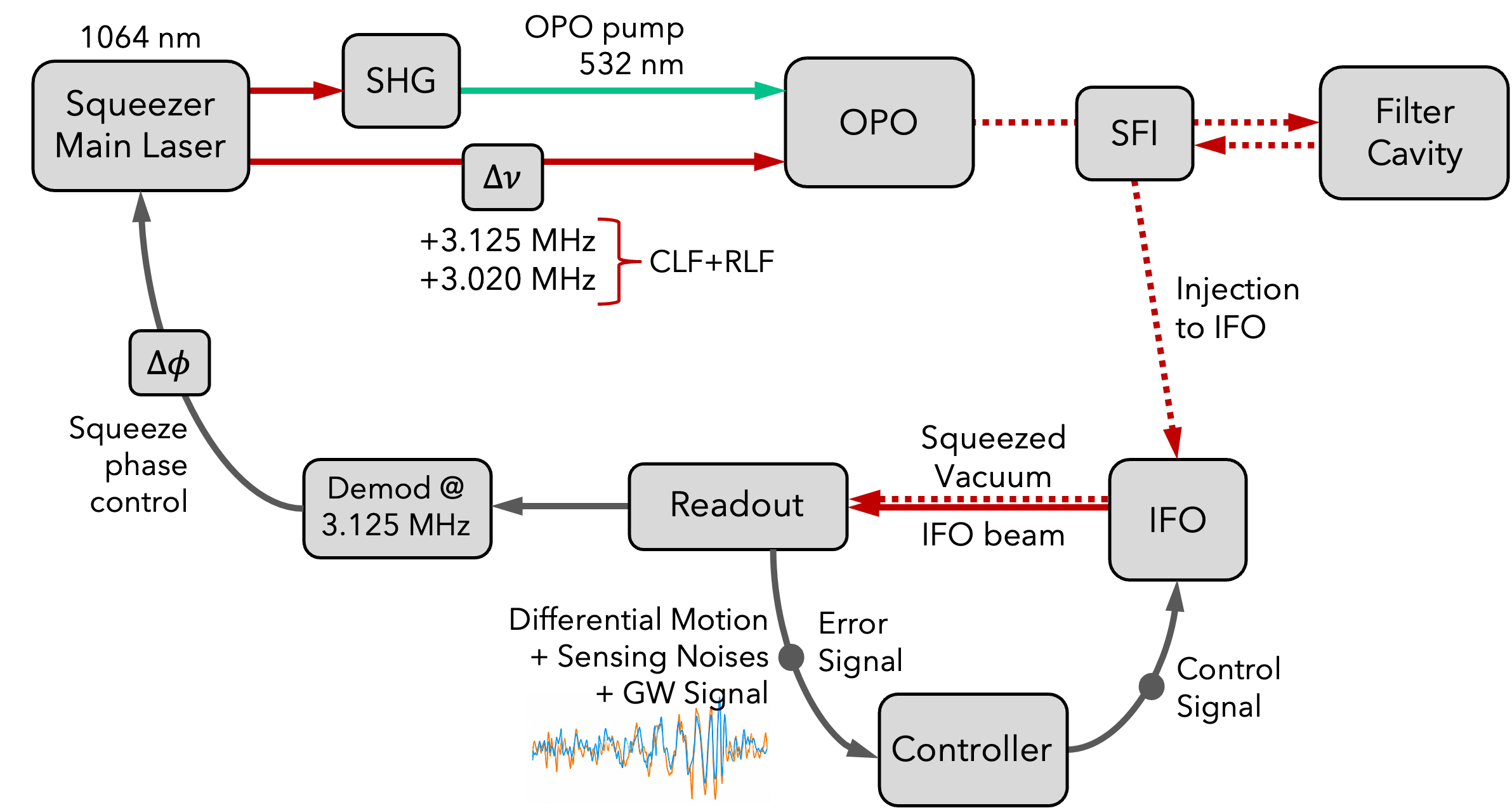}
    \caption{Simplified signal diagram of the squeezing injection and readout system in LIGO. Red and green lines denote optical signals with 1064 nm and 532 nm wavelength respectively. Grey lines denote electrical signals. In Frequency Independent Squeezing configuration, the end mirror of the filter cavity is misaligned, rendering the filter cavity input mirror a simple reflector. The readout signal representation was adapted from the manuscript reporting the first ever detection of gravitational waves~\cite{GW150914}, Abbott et al., Observation of Gravitational Waves from a Binary Black Hole Merger, Phys. Rev. Lett. 116, 061102 (2016), doi:10.1103/PhysRevLett.116.061102, published by the American Physical Society under the Creative Commons Attribution 3.0 License; changes were made.
    }
    \label{fig:sqz_diagram}
\end{figure}

To understand how this is achieved, we outline how the squeezing subsystem works. Figure~\ref{fig:sqz_diagram} shows a simplified diagram. Some portion of the squeezer main laser(at 1064 nm) is sent to the second harmonic generator (SHG)\footnote{Two photons with $\sim$1064 nm wavelength coming from the squeezer main laser are converted to one photon with  $\sim$532 nm wavelength. This 532 nm beam is then used as the pump of the OPO.} to generate the pump beam (532 nm) for the OPO. The squeezer main laser is also used to create the coherent locking field (CLF) and the resonant locking field (RLF)\footnote{Standard cavity locking techniques require some kind of effective sideband~\cite{PDH}, RLF is the sideband for filter cavity locking. The filter cavity was installed as part of the frequency-dependent squeezing upgrade to LIGO, discussed in Section~\ref{sec:fds}}. Nonlinear interactions in the OPO generate squeezed vacuum as well as the conjugate sidebands of the CLF/RLF beams, all of which(dotted red line) are reflected from the filter cavity. The beams going to and returning from the filter cavity are separated with a free-space directional coupler called squeezer Faraday isolator (SFI) and sent to the interferometer. 

The interferometer readout signal contains information about the differential motion of the interferometer test masses, including gravitational-wave signals and residual classical motion, together with sensing noise. This readout also serves as the error signal for the differential-arm control loop. A controller feeds back to the test-mass actuators, suppressing differential arm motion and keeping the interferometer close to its fixed DC readout operating point. The residual error signal, together with the applied control signal, retains the information needed to reconstruct the differential arm motion.

The photodetector readout is also demodulated at the CLF frequency, extracting the beat between the CLF and the interferometer field. Since the CLF experiences the same nonlinear interaction as the squeezed vacuum, its phase is tied to the OPO pump phase and therefore to the orientation of the squeezing ellipse. The 3.125 MHz signal measured at the interferometer readout can therefore be used as a proxy for the relative phase between the squeezing ellipse and the interferometer field~\cite{CLF,Aasi2013Squeezing,Dwyer2013Quadrature,Tse2019QuantumEnhanced}. The resulting phase error signal is fed back to the Squeezer Main Laser, adjusting the squeezer phase relative to the interferometer and stabilising the orientation of the squeezing ellipse.

\subsubsection{The Effect of FIS on Sensitivity}
\label{sec:fis}

The squeezed vacuum states injected through the dark port travel through the interferometer, which transforms the ellipse due to optomechanical effects (see Section~\ref{sec:prop_and_trans} for more on this). Comparing the detector sensitivity with and without squeezed-vacuum injection shows how these changes ultimately affect the measured strain noise.

\begin{figure}[b!]
    \centering
    \includegraphics[width=0.9\linewidth]{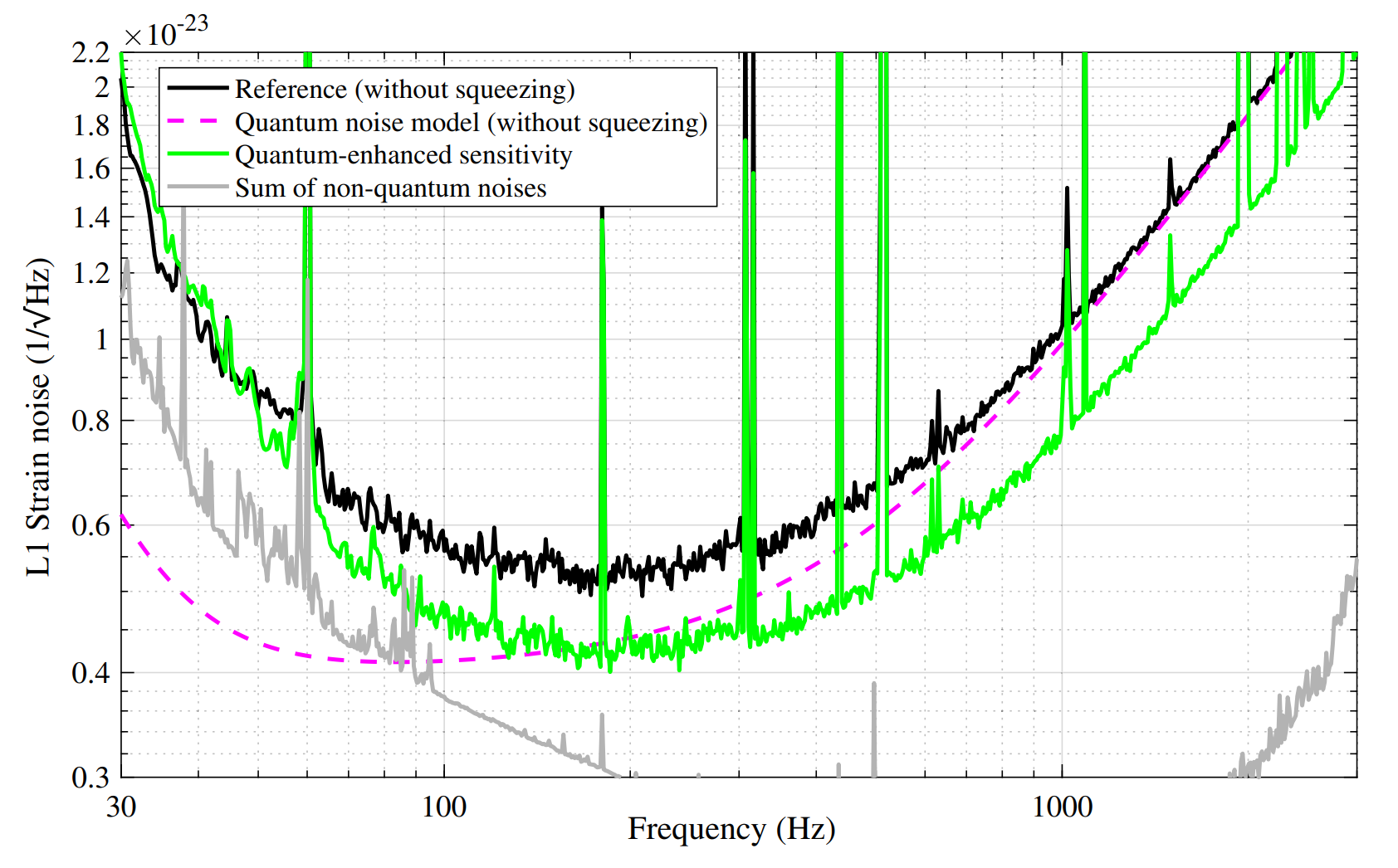}
    \caption{
    Frequency-independent squeezing in the LIGO Livingston detector.
    The black trace is the measured reference sensitivity without squeezing,
    while the green trace shows the measured sensitivity with squeezed-vacuum
    injection. The dashed magenta trace is the modelled quantum noise without
    squeezing, and the grey trace is the estimated sum of the non-quantum
    noise sources. Squeezing improves the measured sensitivity where quantum
    noise is a significant part of the total detector noise. Reproduced from
    Ref.~\cite{Tse2019QuantumEnhanced}, © The Author(s) 2019, licensed under CC BY 4.0.
    }
    \label{fig:fis_tse}
\end{figure}

The results from LIGO implementation of squeezing are shown in Figure~\ref{fig:fis_tse}. The measured sensitivity with
squeezing is lower than the reference (without squeezing) sensitivity over much of the frequency band above
roughly \(50\,{\rm Hz}\), where quantum noise forms a substantial part
of the detector noise. This was the first use of squeezed vacuum states in
Advanced LIGO gravitational-wave observations and produced up to
\(3\,{\rm dB}\) improvement in sensitivity~\cite{Tse2019QuantumEnhanced}.

Notably, below 50 Hz, it appears that the noise floor is slightly increased with injected squeezing. The difference is small enough that it could have been caused by other effects, such as elevated ground motion. Nevertheless, the quantum-noise model predicts precisely such a tradeoff: reducing the phase-like fluctuations responsible for shot-noise imprecision necessarily anti-squeezes the orthogonal amplitude-like quadrature, increasing radiation-pressure back-action. The squeezing angle was chosen primarily to reduce the phase-like fluctuations that limit the high-frequency readout. The orthogonal, amplitude-like quadrature was therefore anti-squeezed. These amplitude fluctuations modulate the optical power circulating in the interferometer and hence the fluctuating radiation-pressure force acting on the test masses. Thus, the same squeezed state that reduces measurement imprecision also increases quantum back-action. This effect is apparent in Fig. \ref{fig:FDS_ganapathy}.

\begin{figure}[hbt!]
    \centering
    \includegraphics[width=0.7\linewidth]{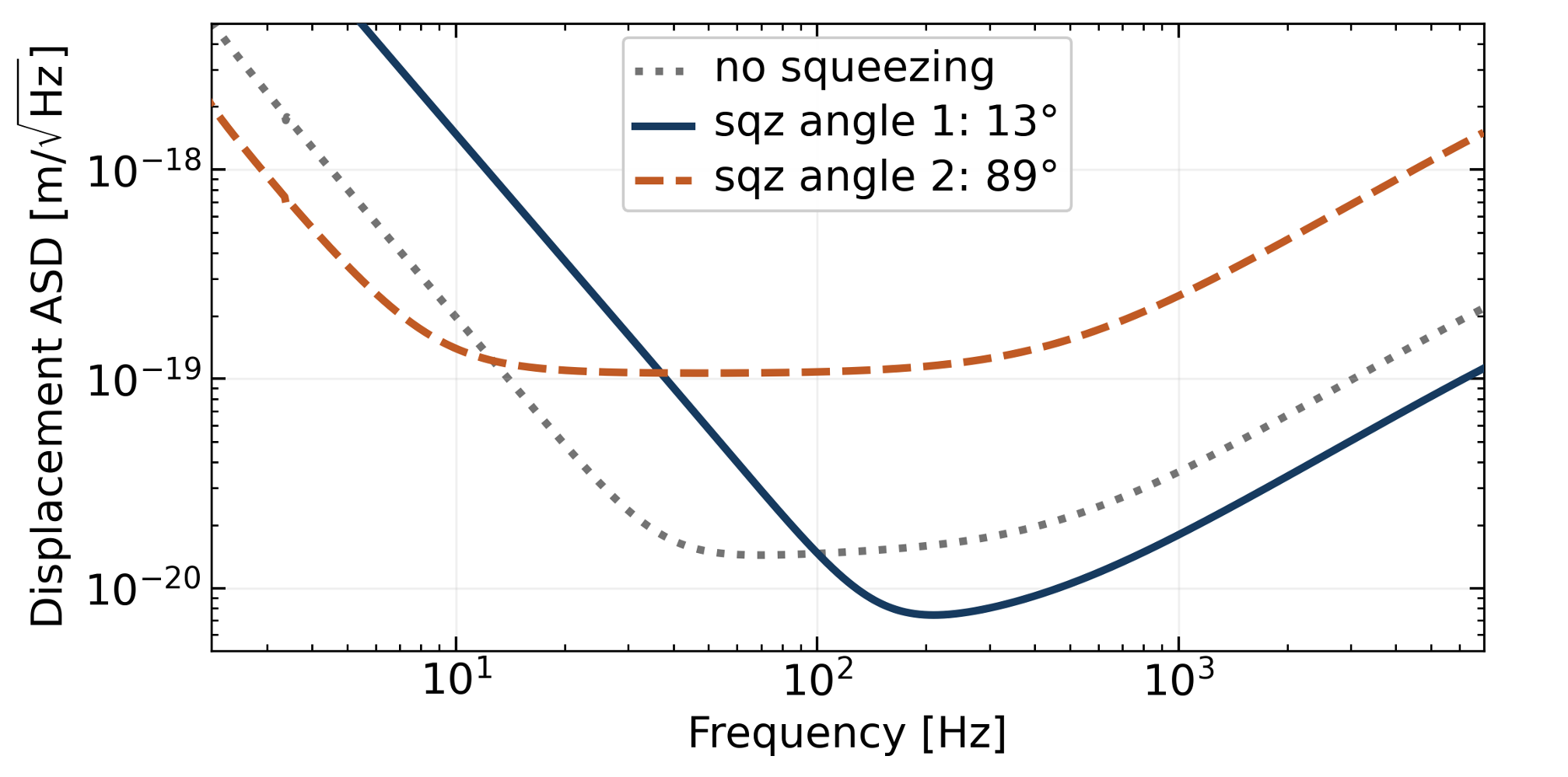}
    \caption{ Modelled quantum displacement noise for two frequency-independent squeezing angles, compared with ordinary vacuum. The near phase-squeezed orientation reduces high-frequency shot-noise imprecision but increases low-frequency radiation-pressure back-action. Rotating the squeezing angle towards the other quadrature exchanges these effects: radiation-pressure noise is reduced while shot noise is increased. No single frequency-independent squeezing angle reduces both contributions across the full detection band.
    }
    \label{fig:fis_sqz_asqz}
\end{figure}

If the squeezing angle is instead rotated such that the amplitude-like quadrature is squeezed while the phase-like quadrature is anti-squeezed, radiation-pressure noise is then reduced, at the expense of increased shot noise, see Figure~\ref{fig:fis_sqz_asqz}. Neither orientation is optimal across the entire detection band: phase squeezing is advantageous at high frequencies, where measurement imprecision dominates, whereas amplitude squeezing is advantageous at low frequencies, where radiation-pressure back-action becomes important.

Essentially, the transformation applied to the squeezing ellipse by the interferometer is frequency dependent, and consequently the squeezing quadrature that minimises the measured quantum noise is also frequency dependent. At high frequencies, the preferred orientation reduces phase-like fluctuations and therefore shot-noise imprecision, while at low frequencies the preferred orientation is rotated towards the orthogonal quadrature so that amplitude-like fluctuations and radiation-pressure back-action are reduced. If the injected squeezing ellipse could be rotated continuously with frequency to follow this changing optimum, squeezing could reduce quantum noise across the full observation band. This is the motivation for \textit{frequency-dependent squeezing}.

\FloatBarrier

\subsection{Frequency-dependent squeezing (FDS)}
\label{sec:fds}

Frequency-dependent squeezing, or FDS, is an input quantum-state engineering technique in which a filter cavity rotates the squeezing angle as a function of frequency before the squeezed field enters the interferometer~\cite{Kimble2001Conversion}. The corresponding input quantum states in the quadrature picture were shown in the right panel of Figure~\ref{fig:input_states}. Audio-band frequency-dependent squeezing was first demonstrated in prototype filter-cavity systems and subsequently developed for implementation in
Advanced LIGO~\cite{Oelker2016AudioBandFDS,McCuller2020FDS}. LIGO integration was completed in time for Observing Run 4 in 2023~\cite{Ganapathy2023Broadband,Jia2024SQL}. 

\begin{figure}[hbt!]
    \centering
    \includegraphics[width=0.65\linewidth]{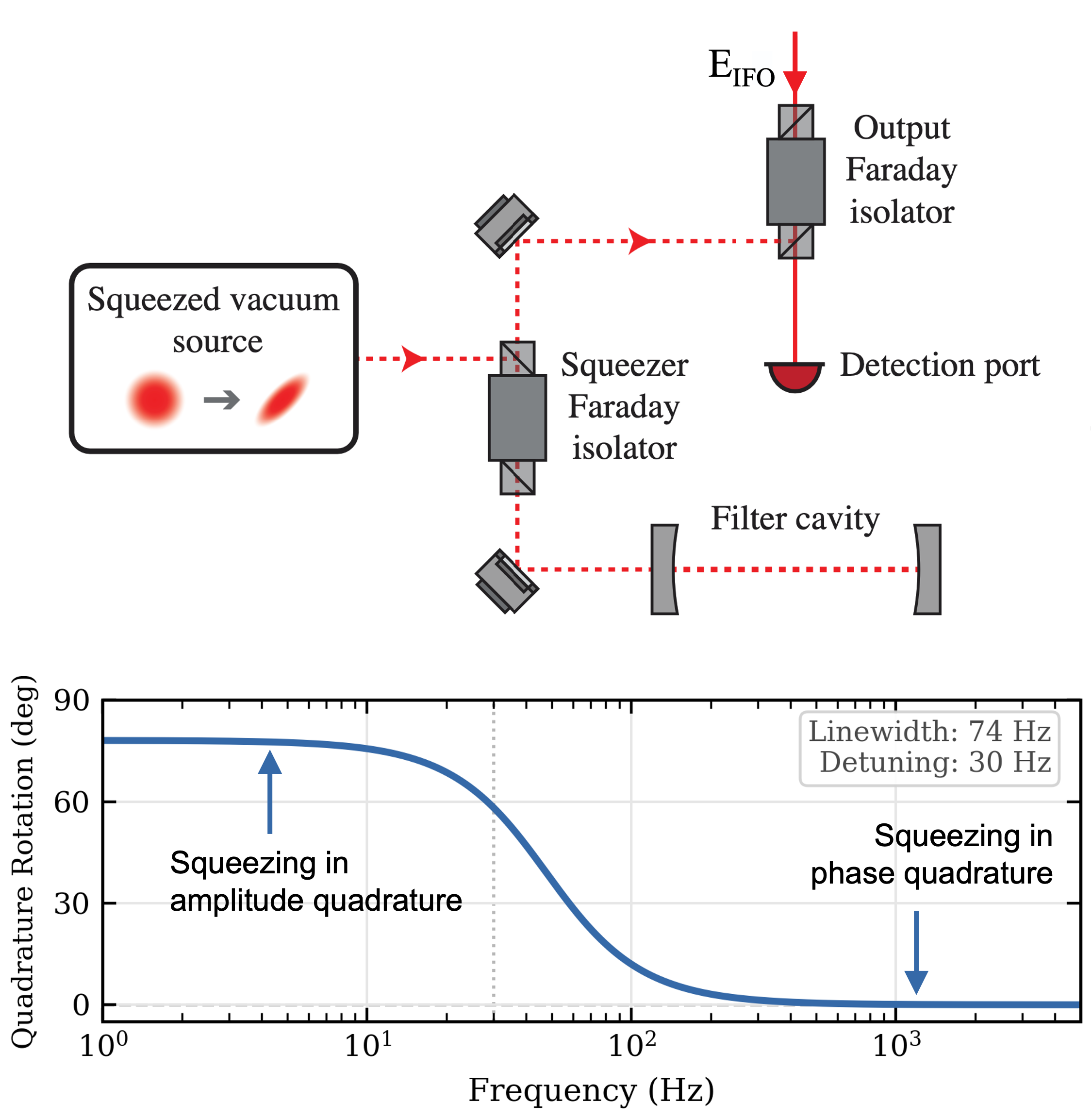}
    \caption{Top panel: Simple diagram for frequency dependent squeezing optical path. Adapted from~\cite{Dwyer2022Squeezing}, © The Author(s) 2022, licensed under CC BY 4.0., changes were made. Bottom panel: Squeezing angle acquired by the squeezed state, i.e. quadrature angle of the squeezing ellipse, after reflection from the filter cavity, as a function of frequency. Filter cavity parameters used for this figure (LIGO Livingston installed filter cavity~\cite{Jia2024SQL}) are indicated on the top right.
    }
    \label{fig:simple_fc_rot}
\end{figure}

The top panel of Figure~\ref{fig:simple_fc_rot} shows a simple diagram of the optical path used to prepare frequency-dependent squeezing. Squeezed vacuum is generated with the OPO, then it is sent to the filter cavity through a Faraday isolator\footnote{Similar to the FIS injection into the interferometer, a Faraday isolator is a convenient, low-loss solution for directional coupling of light.}. The field is reflected from the 300-m-long filter cavity, travels back through the Squeezer Faraday isolator, and couples to the interferometer with the Output Faraday isolator. The long cavity length makes it possible to produce the narrow response required for the quadrature rotation while keeping the round-trip optical loss sufficiently small~\cite{Evans2013RealisticFilterCavities}.

The bottom panel of Figure~\ref{fig:simple_fc_rot} shows the frequency-dependent squeezing angle acquired after reflection from the filter cavity, corresponding to the rotation of the squeezing ellipse. The cavity linewidth and detuning determine the frequency range over which the phase transition occurs. Because the upper and lower audio sidebands acquire frequency-dependent phase shifts upon reflection, their combined phase is modified, which appears as a rotation of the squeezing ellipse in quadrature space. The cavity linewidth and detuning are chosen so that the filter-cavity rotation approximately compensates the frequency-dependent transformation produced by the interferometer~\cite{Kimble2001Conversion,
Ganapathy2023Broadband,McCuller2021QuantumResponse}. 

\begin{figure}[hbt!]
    \centering
    \includegraphics[width=\linewidth]{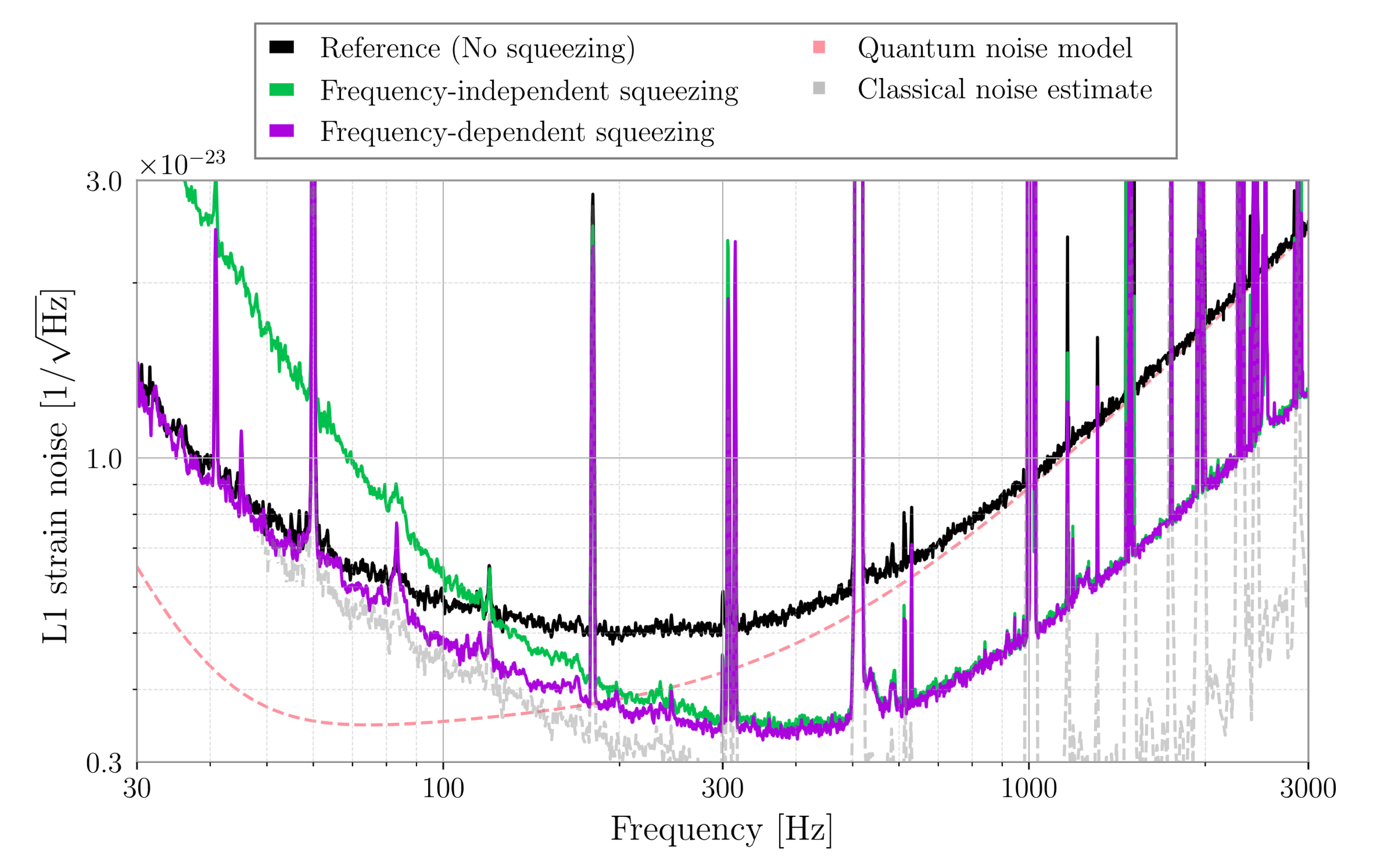}
    \caption{Measured LIGO sensitivity without squeezing, with frequency-independent squeezing, and with frequency-dependent squeezing. Frequency-independent squeezing reduces shot noise at high frequencies but increases radiation-pressure noise at low frequencies. Reflection from the filter cavity rotates the squeezing quadrature with frequency, retaining the high-frequency improvement while recovering the low-frequency sensitivity. Reproduced from Ref.~\cite{Ganapathy2023Broadband}, © The Author(s) 2023, licensed under CC BY 4.0.}
    \label{fig:FDS_ganapathy}
\end{figure}

In the measurements preceding Observing Run 4, shown in Figure~\ref{fig:FDS_ganapathy}, squeezing reduced the shot-noise amplitude around 1 kHz by $\sim$4 dB at LIGO Hanford and $\sim$5.8 dB at LIGO Livingston, while frequency-dependent squeezing additionally reduced the total detector noise by approximately 1-2 dB between 60 and 100 Hz~\cite{Ganapathy2023Broadband}. This was the first broadband quantum enhancement of full-scale gravitational-wave detectors, extending from tens of hertz to several kilohertz.

This result demonstrates the basic purpose of the filter-cavity rotation: different parts of the detection band require different orientations of the injected squeezing ellipse. We have already motivated this need by considering the two limiting regimes: at low frequencies, where radiation-pressure noise is important, and at high frequencies, where shot noise dominates. The effect is more subtle for frequencies in between these two limits.

During the transition from phase squeezing to amplitude squeezing, the squeezing ellipse is tilted with respect to the amplitude and phase quadratures, see Figure~\ref{fig:input_states} right panel. A tilted ellipse means that fluctuations in these two quadratures are correlated. This can be described by their covariance. If fluctuations in the two quadratures are uncorrelated, their covariance is zero; if a fluctuation in one quadrature tends to occur together with a fluctuation of a particular sign in the other, the covariance is nonzero. Geometrically, an ellipse whose principal axes are aligned with the amplitude and phase axes has zero covariance between those quadratures, while a tilted ellipse has nonzero covariance. This means, the $S_{xx}^{imp,ba}$ term in Equation~\ref{eq:qnoise_decomposition} becomes decidedly nonzero. 

Frequency-dependent squeezing therefore does more than smoothly interpolate between phase squeezing and amplitude squeezing. At intermediate frequencies, the filter cavity intentionally prepares an input quantum state with correlated amplitude and phase fluctuations. These correlations are important because the two quadratures play different roles in the interferometric measurement: amplitude-like fluctuations drive radiation-pressure back-action, while phase-like fluctuations contribute to measurement imprecision.

The squeezing ellipse prepared at the interferometer input, however, is not the ellipse that reaches the readout. The optical and optomechanical response of the interferometer further transforms the quantum state, changing both the fluctuations in the individual quadratures and the correlations between them. The optimum filter-cavity rotation must therefore be chosen with these transformations in mind.

To understand what the filter cavity is compensating, why the required squeezing angle varies continuously with frequency, and how the resulting correlations can affect the total quantum noise, we first need to follow the propagation and transformation of the quantum state through the interferometer.

\FloatBarrier

\section{From Injected Squeezing to Measured Quantum Noise}

\subsection{Transformation of the Quantum State through the Interferometer}
\label{sec:prop_and_trans}
Let us first consider what happens to ordinary vacuum as it propagates through the interferometer. As mentioned before, at the input, ordinary vacuum is represented by a circular noise distribution: its amplitude and phase quadratures have equal variance and zero covariance. If the interferometer simply transmitted these fluctuations without mixing them, the state would remain circular at the readout. This is not the case.

Recall that LIGO is an optomechanical system: the optical field interacts with movable mirrors. An amplitude fluctuation in the field changes the optical power circulating in the arm cavities and, therefore, changes the radiation-pressure force acting on the test masses. The mirrors move in response to this force, and their displacement changes the phase of the light leaving the interferometer. An input amplitude fluctuation can therefore appear in the outgoing field through two paths: directly as an amplitude fluctuation, and indirectly as a phase fluctuation generated through radiation-pressure-driven mirror motion.

\begin{figure}[hbt!]
    \centering
    \includegraphics[width=0.95\linewidth]{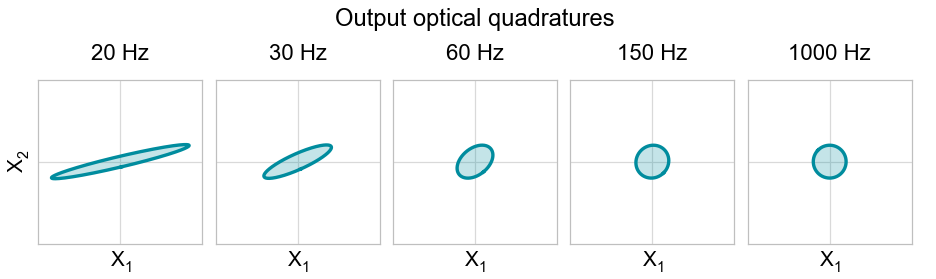}
    \caption{Output optical-quadrature distributions for ordinary-vacuum input at representative frequencies. The full interferometer model is used to compute these ellipses~\cite{McCuller2021QuantumResponse, GWINC}. Radiation-pressure coupling between the optical field and the test masses shears the initially circular vacuum distribution, producing a tilted ellipse and correlations between the output quadratures. The optomechanical transformation is strongest at low frequencies and becomes progressively weaker towards high frequencies. 
    }
    \label{fig:shearing_nosqz}
\end{figure}

The outgoing amplitude and phase quadratures are therefore no longer necessarily uncorrelated. In the ellipse picture, the initially circular vacuum state is sheared into a tilted ellipse, as illustrated in Figure~\ref{fig:shearing_nosqz}. This optomechanically generated correlation between the optical quadratures is commonly called \textit{ponderomotive squeezing}~\cite{Braginsky1967Ponderomotive}. Importantly, this effect is present with or without an injected squeezed state: the correlation is generated by the interaction between the quantum fluctuations of the light and the motion of the mirrors.

LIGO of course contains more than the optomechanical interaction with the arm-cavity mirrors. The quantum field propagates through a coupled system of optical cavities before reaching the readout, and these optical elements can further transform the quadratures. As we saw for the filter cavity, a detuned cavity can impart different phase shifts to the upper and lower audio-frequency sidebands and therefore rotate the quadrature ellipse. The signal-recycling cavity and the arm cavities together determine the frequency-dependent optical response of the interferometer, while detuning can introduce additional quadrature rotation. For an already squeezed state, such rotations change the orientation of the ellipse relative to the amplitude and phase axes and therefore change their covariance. A pure rotation alone, however, cannot produce correlations from ordinary vacuum: a circle remains a circle under rotation.

\begin{figure}[hbt]
    \centering
    \includegraphics[width=0.8\linewidth]{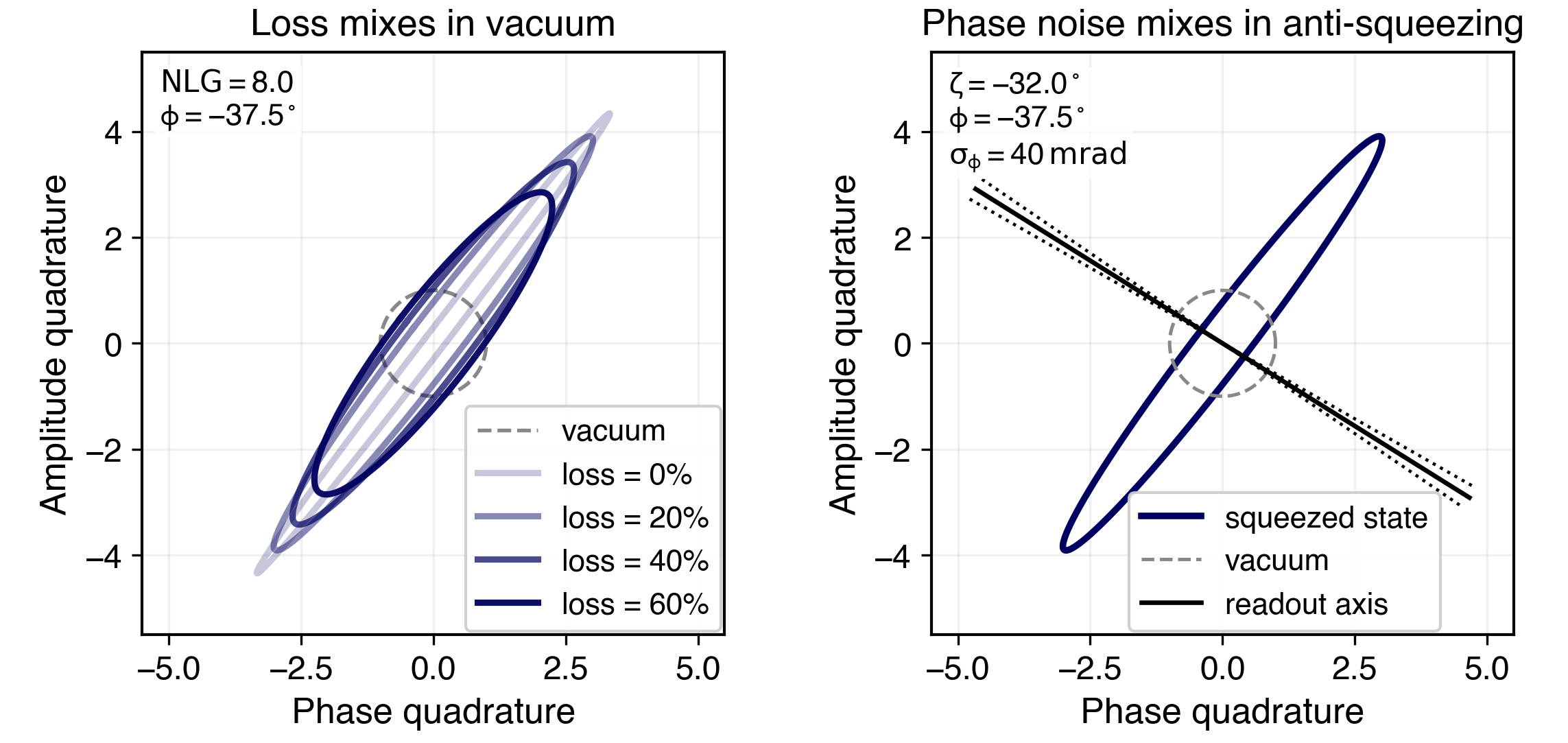}
    \caption{
    Effects of optical loss and squeeze-angle noise on a 14.8 dB squeezed state.
    Optical loss replaces part of the prepared state with ordinary vacuum,
    rounding the squeezing ellipse. Phase noise ($\sigma_\phi$) makes the ellipse orientation (squeezing angle, $\phi$)
    fluctuate relative to the readout axis (readout angle, $\zeta$) and therefore couples part of the
    anti-squeezed quadrature into the measurement.
    }
    \label{fig:squeezing_loss_phase_noise}
\end{figure}

Not all transformations preserve the quantum state. At every lossy interface, some fraction of the field is lost and ordinary vacuum enters in its place. For ordinary vacuum this makes no observable difference, since vacuum mixed with vacuum remains vacuum. For a squeezed state, however, the incoming vacuum has larger fluctuations along the squeezed direction than the state it replaces. The squeezed axis therefore broadens and the ellipse is pulled back towards the ordinary-vacuum circle (Figure~\ref{fig:squeezing_loss_phase_noise} left panel). At the photon-pair level, this degradation can also be understood as a loss of the correlations responsible for squeezing. If one member of a correlated pair is lost, that pair can no longer contribute its full correlation to the detected state. The lost portion of the field is effectively replaced by uncorrelated vacuum fluctuations. As a result, optical loss reduces the observable squeezing and, more generally, reduces our ability to preserve the engineered covariance of the input state~\cite{Evans2013RealisticFilterCavities}.

Another important degradation is phase noise. The benefit of squeezing depends on maintaining the orientation of the squeezing ellipse relative to the quadrature required by the interferometer. Residual fluctuations in this relative angle cause the ellipse to jitter around its desired orientation (Figure~\ref{fig:squeezing_loss_phase_noise} right panel). The detector therefore samples a small amount of the anti-squeezed quadrature together with the squeezed quadrature. When averaged over time, this increases the measured noise along the nominally squeezed direction. The effect becomes increasingly important as larger squeezing levels are generated, because stronger squeezing is accompanied by stronger anti-squeezing and even a small angular fluctuation can couple appreciable anti-squeezed noise into the measurement.

\FloatBarrier

Therefore, we can think of the quantum state as being continuously reshaped as it propagates through the detector. The optomechanical interaction shears the ellipse and generates correlations, optical cavities rotate it, loss broadens it, and phase noise causes its orientation to fluctuate. The state that finally reaches the readout is the result of all of these transformations acting on the state originally injected into the interferometer. This transformation can be expressed in a compact form using a multiport input--output description~\cite{Kabagoz2026BackAction, McCuller2021QuantumResponse},
\begin{equation}
    \hat{\mathbf b}(\Omega)
    =
    \mathbf T_{\rm sqz}(\Omega)\hat{\mathbf a}_{\rm sqz}(\Omega)
    +
    \sum_k \mathbf T_k(\Omega)\hat{\mathbf n}_k(\Omega),
\end{equation}
where \(\hat{\mathbf b}\) is the two-quadrature field at the interferometer readout, \(\hat{\mathbf a}_{\rm sqz}\) is the squeezed field injected through the dark port, and \(\hat{\mathbf n}_k\) represents the vacuum fields entering through the other optical and loss ports. The frequency-dependent transfer matrices \(\mathbf T_{\rm sqz}\) and \(\mathbf T_k\) describe how each of these fields is transformed by the optical and optomechanical response of the interferometer before reaching the readout. In this picture, the quantum state at the readout is the combined result of the injected squeezed state, the interferometer transformation, and the additional vacuum fields introduced through loss.

\subsection{From Quadrature Correlations to Quantum Noise}

What we ultimately measure is a projection of this output quantum state onto the interferometer readout quadrature, $\hat b_{\zeta}(\Omega)=\hat b_1(\Omega)\cos\zeta+\hat b_2(\Omega)\sin\zeta$,
where \(\hat b_1\) and \(\hat b_2\) are the two orthogonal output-field quadratures and $\zeta$ is the readout quadrature angle. At each frequency, the variance of the ellipse along this readout direction determines the quantum noise in the detected optical signal. Once calibrated by the interferometer response, this becomes the equivalent displacement-noise spectrum $S_{xx}^{\rm QN}$ that appears in the detector sensitivity. When the output optical fluctuations are referred back to an equivalent test-mass displacement, the quadrature correlations described above can appear as correlations between measurement imprecision and radiation-pressure back-action. The quantum noise measured at the detector can therefore be decomposed into three contributions: measurement imprecision, quantum back-action, and the correlation between them, see Equation~\ref{eq:qnoise_decomposition}.

 The first two terms are noise power spectral densities and are necessarily non-negative. The correlation term, in contrast, can have either sign. The quantum noise components are subject to the Heisenberg uncertainty principle shown below for symmetrised, single-sided power spectral densities~\cite{Clerk2010QuantumNoise,BraginskyKhalili1992}: 

\begin{equation}
    S_{xx}^{\rm imp}S_{xx}^{\rm ba} - \frac{|S_{xx}^{\rm imp,ba}|^2}{4} \geq {\hbar^2|\chi_m|^2.}
    \label{eq:heisenberg}
\end{equation}

\noindent In the observation band, well above their natural pendulum frequency of $\sim0.5$ Hz, the suspended test masses behave approximately as free masses, for which $|\chi_m(\Omega)| \propto 1/{\Omega^2}$. This is why a given radiation-pressure force fluctuation produces increasingly large displacement noise towards low frequencies (see Equation~\ref{eq:displacement_referred_terms}).

If no correlations are present, Equation~\ref{eq:heisenberg} simplifies to the familiar measurement trade-off: reducing measurement imprecision requires increased uncertainty in the back-action force, and balancing these two contributions gives the conventional standard quantum limit. On the other hand, for a given nonzero correlation, the uncertainty relation requires a correspondingly larger minimum product of the imprecision and back-action uncertainties.
Yet, the detector sensitivity only sees the total quantum noise. Therefore, with negative correlations, the total quantum noise can decrease even though the individual imprecision and back-action contributions may increase.

\begin{figure}[hbt!]
    \centering
    \includegraphics[width=\linewidth]{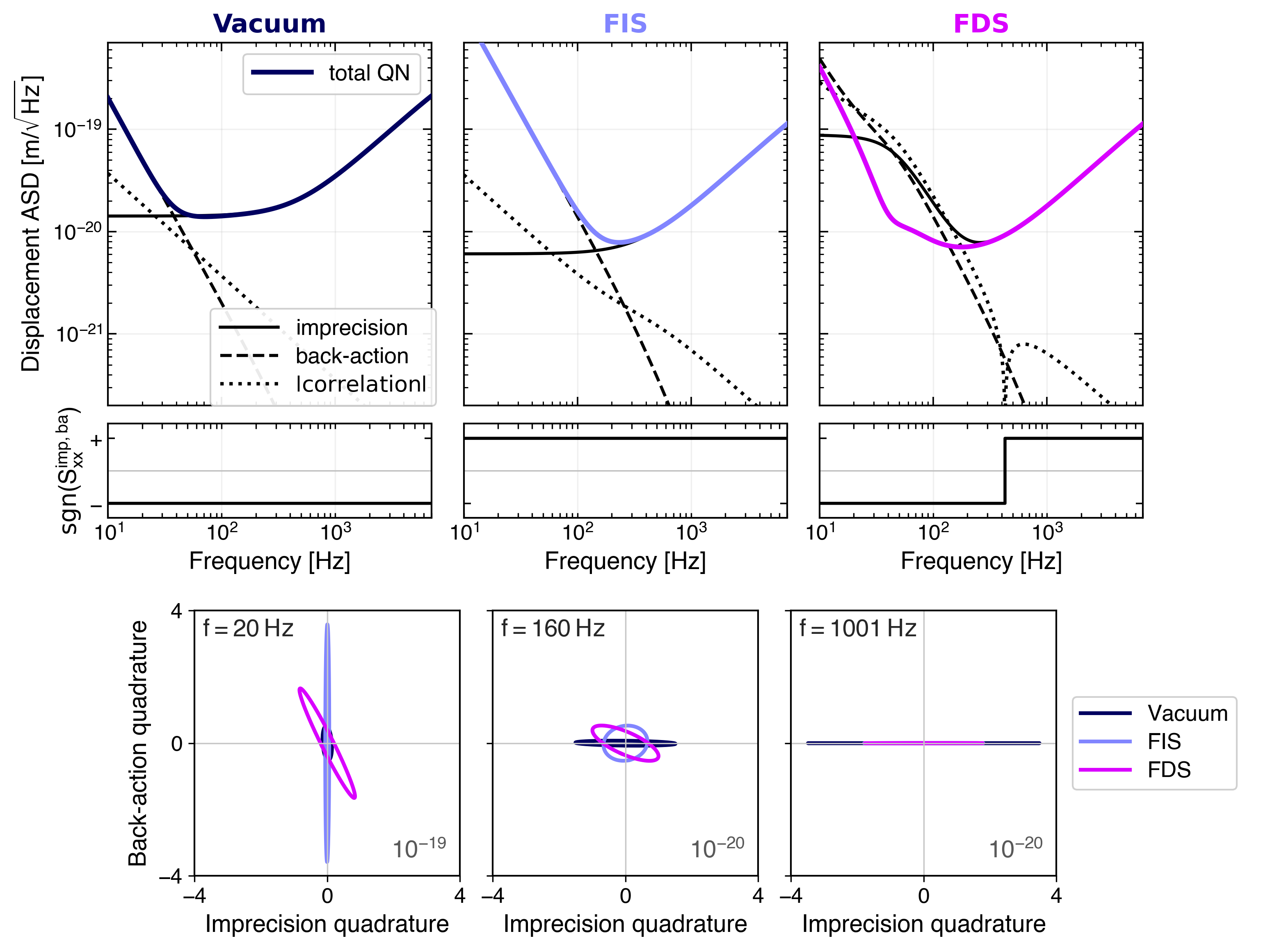}
    \caption{
    Decomposition of the modelled quantum noise for ordinary vacuum, frequency-independent squeezing (FIS), and frequency-dependent squeezing (FDS) into measurement imprecision, radiation-pressure back-action, and their correlation in amplitude spectral density units ($\sqrt{S_{xx}^{imp}}, \sqrt{S_{xx}^{ba}}, \sqrt{|S_{xx}^{imp,ba}|}$). The decomposition in Equation~\ref{eq:qnoise_decomposition} applies to the corresponding power spectral densities. Because the correlation contribution can have either sign, its magnitude is shown in the upper panels and its sign separately below. The lower panels show the corresponding output-referred displacement-noise covariance ellipses at
    representative frequencies, illustrating how the frequency-dependent quadrature fluctuations map onto the measured quantum-noise contributions. 
    }
    \label{fig:decomposed_ellipses}
\end{figure}

Numerical models of the full quantum-noise response of complex interferometers
such as LIGO already exist~\cite{McCuller2021QuantumResponse,GWINC}. These
models account for the coupled optical and optomechanical response of the
detector and predict the total quantum noise at the readout. The same numerical
models can also be leveraged to apportion this total into the individual
measurement-imprecision, quantum-back-action, and correlation contributions
introduced above~\cite{Kabagoz2026BackAction}.

Figure~\ref{fig:decomposed_ellipses} shows this decomposition for ordinary vacuum, frequency independent squeezing, and frequency dependent squeezing. Because the correlation contribution to the quantum-noise power spectral density can be either positive or negative, the upper panels show $\sqrt{|S_{xx}^{\rm imp,ba}|}$ in amplitude spectral density units, while its sign is shown separately below. The lower panels show the corresponding quantum-state ellipses at representative frequencies. The decomposition is particularly useful in the intermediate-frequency region, where measurement imprecision and radiation-pressure back-action are both appreciable and where the detector is especially sensitive to compact-binary signals, including binary-neutron-star inspirals.

The decomposition also shows why the low-frequency effect of FIS is more subtle than simply an increase in radiation-pressure noise. The squeezed
input changes both the back-action contribution and its correlation with measurement imprecision. For FIS, this correlation contributes with the
unfavourable sign, further increasing the total quantum noise. With FDS, the frequency-dependent input covariance instead produces a negative correlation over the intermediate-frequency region, allowing it to cancel part of the positive imprecision and back-action contributions. The intermediate squeezing angles therefore do more than provide a smooth transition between amplitude and phase squeezing: after propagation through the interferometer, their covariance can contribute directly to reducing the measured quantum noise.

This brings us back to the standard quantum limit introduced in Section~\ref{sec:shot-rad}. The conventional SQL is obtained by optimising the trade-off between measurement imprecision and quantum back-action when the two contributions are uncorrelated. It is therefore an important benchmark for continuous quantum measurements, but it is not a fundamental
lower bound once correlations are allowed. As Equation~\ref{eq:heisenberg} shows, the imprecision, back-action, and correlation terms remain constrained
by the uncertainty principle. Nevertheless, when
$S_{xx}^{\rm imp,ba}<0$, the correlation term subtracts from the two positive noise contributions and allows the total measured quantum noise to fall
below the uncorrelated SQL.

\begin{figure}[hbt!]
    \centering
    \includegraphics[width=\linewidth]{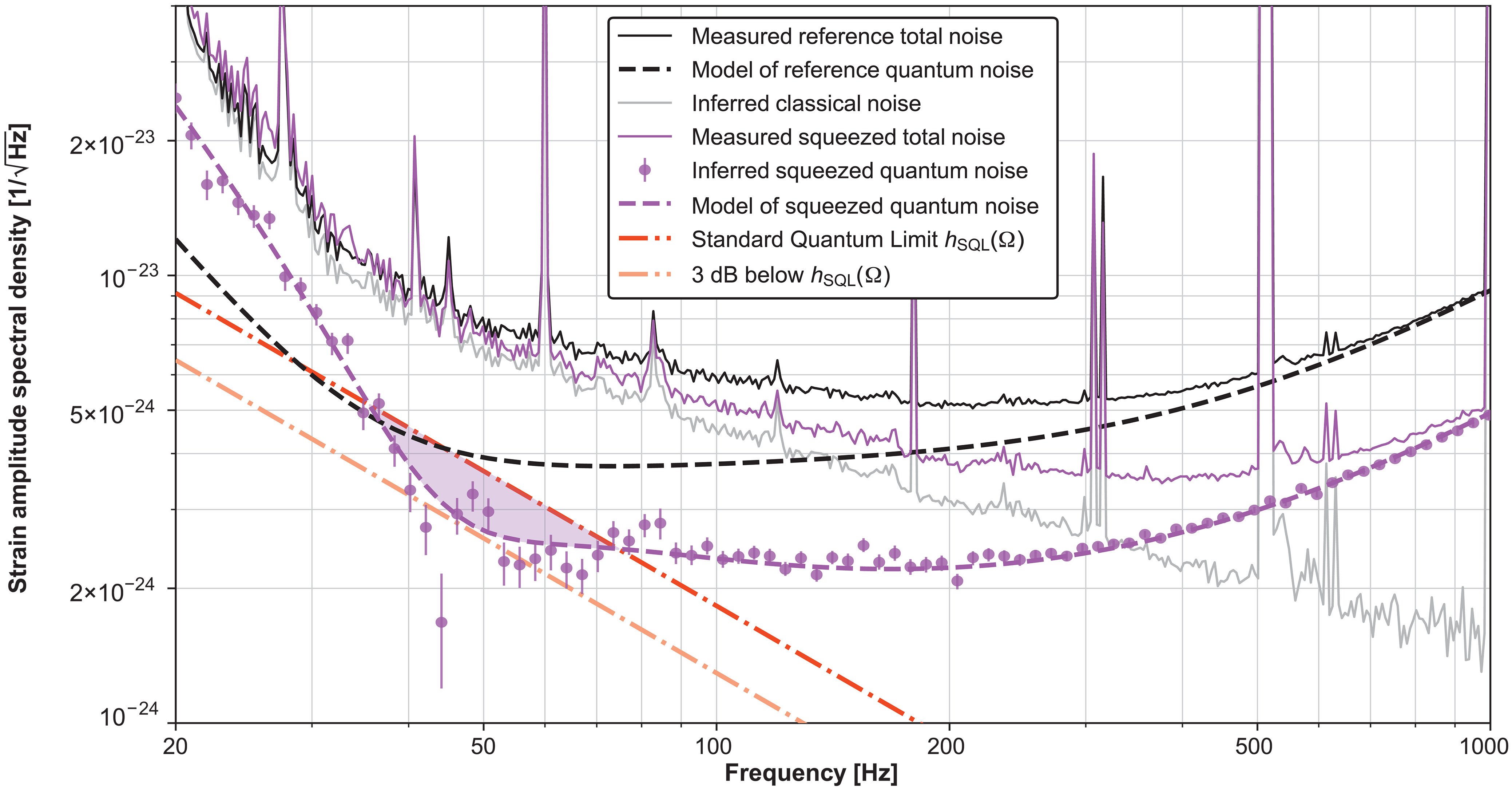}
    \caption{
    Measured quantum noise at LIGO Livingston compared with the free-mass
    standard quantum limit. Frequency-dependent squeezing reduces the quantum
    noise below the SQL over part of the low-frequency observation band. Adapted from Fig.~2 of the \href{https://arxiv.org/abs/2404.14569} {accepted author manuscript} of Ref.~\cite{Jia2024SQL}. \textcopyright~2024 The Authors, licensed under \href{https://creativecommons.org/licenses/by/4.0/}{CC BY 4.0}.
    }
    \label{fig:sql_jia}
\end{figure}

Frequency-dependent squeezing has allowed this regime to be reached experimentally in LIGO. Figure~\ref{fig:sql_jia} shows the measured quantum
noise at LIGO Livingston relative to the free-mass standard quantum limit. With frequency-dependent squeezing, the quantum noise was reduced below the
SQL between approximately $35$ and $75\,{\rm Hz}$, reaching nearly $3\,{\rm dB}$ below the SQL around $50\,{\rm Hz}$~\cite{Jia2024SQL}. This provides a direct experimental demonstration that correlations between measurement imprecision and quantum back-action can be used to surpass the conventional standard quantum limit without violating the uncertainty principle.

\newpage

\section{Looking to the Future}
\subsection{Practical limits to observed squeezing}
\label{sec:observed_squeezing_limits}

As discussed in Section~\ref{sec:prop_and_trans}, optical loss and phase noise degrade the squeezed state between its generation
and detection. These effects explain why the squeezing observed at the readout is smaller than the squeezing generated with the OPO.
Therefore, further improvement depends on preserving the prepared state through the full detector, rather than generating more squeezing.

Direct optical losses arise from the injection optics, Faraday isolators, filter cavity, output optics, output mode cleaner, and photodetectors. Mode mismatch at cavity interfaces is a form of optical loss: only the portion of the squeezed field that overlaps the detected interferometer mode contributes to the measurement. For FDS, good mode-matching must be maintained to the filter
cavity, to the interferometer, and finally with the spatial mode selected by the output mode cleaner.

The filter-cavity rotation must also remain matched to the optomechanical response of the interferometer. The optimal rotation depends on the detector configuration, one particularly important parameter is the circulating arm power. Increasing the arm power boosts radiation-pressure coupling and shifts the frequency at which the detector transitions between back-action- and imprecision-dominated
behaviour. The optimal filter-cavity detuning and linewidth therefore change as well. Since the cavity linewidth is largely fixed by its optical design, changes in arm power or other interferometer parameters can leave the
rotation imperfectly matched to the detector, jeopardising the broadband benefit.

The mode matching conditions are also not perfectly static. The beam parameters of the interferometer change with alignment, thermal state, and other operating condition. Advanced LIGO's thermal compensation system (TCS) is designed to counteract thermally induced changes in the interferometer optics and therefore affects the spatial mode of
the interferometer field. Because the squeezed field must overlap this mode, changes in TCS settings, alignment, or beam position can alter the observed squeezing even when the squeezed-light source remains stable.

The observed squeezing is consequently a property of the full detector configuration. Reaching higher squeezing levels at the readout requires reducing direct optical loss, maintaining alignment and mode matching across the squeezing path, and keeping the filter-cavity rotation matched to the evolving interferometer response. To give an idea about the loss levels required for future squeezing goals, with the current phase noise conditions ($\sim20\, \rm mrad$) and assuming the $\sim$18 dB currently generated at the OPO, 10 dB at the readout requires the total loss from OPO to photodiode to be $\sim 6\%$. Today, we are able to obtain $\sim$6 dB squeezing with roughly 23\% total loss~\cite{Kabagoz2025APS,advanced_capote_2025}.

\subsection{Squeezing in future detectors}
\label{sec:sqz_future}

The techniques described here are not the end of the story. The A+ upgrade now underway aims for higher circulating power and lower optical loss~\cite{LSCInstrumentWhitePaper}. The installed filter cavity was originally designed for the A+ goal of 750 kW arm power ($\sim$285 kW today), so the performance below 100 Hz may be improved even further. As mentioned in Section~\ref{sec:background}, for quantum noise reduction to be observable and effective, classical noise contribution should not bury quantum noise. With the lower coating thermal noise planned for the A+ stage, we would be able to leverage quantum noise reduction with frequency dependent squeezing further.  

Beyond A+, proposals such as A$\sharp$ and the third-generation detectors, Cosmic Explorer and the Einstein Telescope, aim higher: heavier test masses and longer arms change the landscape, changing the optimal filter cavity parameters, and the observed squeezing goal is further increased to 10 dB~\cite{LSCInstrumentWhitePaper}. Which design is adopted will depend less on how much squeezing can be generated and more on how much of it can be delivered, which returns us to the loss and mode-matching arguments above. Engineering the input quantum state can be further complemented with readout engineering. 

\subsection{Balanced homodyne detection and macroscopic quantum measurement}
\label{sec:bhd_future}

The discussion so far has focused on engineering the quantum state entering the interferometer. The OPO produces the squeezing ellipse, the filter cavity prepares its frequency-dependent orientation, and the interferometer transforms it through its optical and optomechanical response. In the current DC readout scheme, however, the measured quadrature is fixed.

Balanced homodyne detection(BHD) is planned as a future upgrade to the LIGO readout~\cite{BHDDesign}. Instead of relying on the small carrier field of the DC readout scheme, BHD will mix-in a separate local-oscillator field whose phase can be controlled independently, thereby allowing the arbitrary selection of the readout quadrature.

So, BHD is a natural complement to frequency-dependent squeezing. FDS controls the state injected into the interferometer, while BHD controls how the transformed output state is measured. Selecting the output quadrature provides another experimental handle on the quantum correlations generated in the optomechanical measurement. This creates an opportunity for macroscopic quantum measurement science. 

\section{Closing Remarks}
\label{sec:conclusions}

We have shown that quantum noise in interferometric GW detectors arises from the dual role of light as both the probe of mirror motion and a source of force on the mirrors. Phase-like fluctuations limit the precision of the optical readout, while amplitude-like fluctuations
produce radiation-pressure back-action. These are not unrelated detector noises, but two aspects of the same quantum measurement process, and still subject to the Heisenberg uncertainty principle. 

Squeezed-light injection reduces quantum noise by engineering the input quantum state. A subthreshold OPO correlates optical sidebands, represented in the
quadrature plane by a squeezing ellipse. Frequency-independent squeezing reduces high-frequency measurement imprecision, but is hampered by the increased low-frequency back-action. Frequency-dependent squeezing overcomes this tradeoff by reflecting the squeezed state from a detuned filter cavity before injection.
The resulting frequency-dependent ellipse orientation is better matched to the optomechanical response of the interferometer and produces imprecision--back-action correlations. These correlations can then reduce the total quantum noise, without violating the uncertainty principle. Consequently, frequency-dependent squeezing has enabled broadband quantum-noise reduction in the LIGO detectors, even allowing the quantum noise to dip below the standard quantum limit.

The squeezing observed at the readout is nevertheless a property of the full detector. Optical loss mixes ordinary vacuum into
the prepared state, mode mismatch reduces its spatial overlap with the detected interferometer field, and phase noise and filter-cavity imperfections prevent the intended quadrature from reaching the readout. Further improvement therefore requires
better preservation of the state through the entire optical path it experiences.

Balanced homodyne detection, planned as a future LIGO readout upgrade, adds control of the output measurement basis. With squeezed-state generation at the input, frequency-dependent rotation before injection, optomechanical transformation inside the interferometer, and a tunable quadrature measurement at the output, much of the quantum measurement chain becomes tunable. LIGO can therefore serve not only as a GW observatory, but as a kilometre-scale platform for studying quantum measurement in a macroscopic optomechanical system.

\section*{Acknowledgements}

This material is based upon work supported by NSF's LIGO Laboratory which is a major facility fully funded by the National Science Foundation.\newline

\noindent The author thanks Nergis Mavalvala and Anders Lossius for invaluable discussions. \newline

\noindent OpenAI GPT-5.5 Thinking was used to assist with developing and editing the user interfaces of visualisation scripts available in the LIGO Quantum Tools repository~\cite{Kabagoz2026LIGOQuantumTools}. All scientific content, model choices, figures, and interpretations are the author's own. The author reviewed and verified all AI-assisted output, and no AI tool was used to generate or modify research results.

\section*{Funding}
The author was partially funded by NSF PHY-1834382 and by the European Union (ERC, GRAVITES, No. 101071779).
The author gratefully acknowledges the support of the US National Science Foundation (NSF) for the construction and operation of the LIGO
Laboratory and Advanced LIGO as well as the Science and Technology Facilities Council (STFC) of the United Kingdom, and the Max-Planck-Society (MPS) for support of the construction of Advanced LIGO.
Additional support for Advanced LIGO was provided by the Australian Research Council. LIGO was
constructed by the California Institute of Technology and Massachusetts Institute of Technology with funding from NSF and
operates under cooperative agreement PHY-2309200. Advanced LIGO was built under award PHY-0823459. The A+ upgrade to
Advanced LIGO is supported by US NSF award PHY-1834382 and UK STFC award ST/S00246/1, with additional support from the Australian Research Council. 

\section*{Disclosures}
The author declares no potential conflict of interest.

\section*{Data Availability}
The data used in this work are available upon reasonable request. Open-source visualisation and modelling tools used for the squeezed-state representations and simple squeezing calculations presented in this article are available in the LIGO Quantum Tools repository~\cite{Kabagoz2026LIGOQuantumTools}.

\bibliographystyle{tfnlm}
\bibliography{references}

\end{document}